\documentclass[apj,onecolumn]{emulateapj}
\input epsf.tex
\begin{document}
\title{In Situ Models for Planet Assembly around Cool stars} 
\author{Brad M. S. Hansen
}
\altaffiltext{1}{Department of Physics \& Astronomy, University of California Los Angeles, Los Angeles, CA 90095, hansen@astro.ucla.edu}


\lefthead{Hansen}
\righthead{Cool Star Rocks}

\begin{abstract}
We present a model for the in situ assembly of planetary systems around a 0.5~$M_{\odot}$ star, and compare
the resulting statistics to the observed sample of cool Kepler planet candidates from Dressing \& Charbonneau (2013).
We are able to reproduce the distribution of planetary periods and period ratios, although we once again find
an underabundance of single transit systems relative to the observations. We also demonstrate that almost every
planetary system assembled in this fashion contains at least one planet in the habitable zone, and that water
delivery to these planets can potentially produce a water content comparable to that of Earth. Our results broadly
support the notion that habitable planets are plentiful around M~dwarfs in the solar neighbourhood.

\end{abstract}

\keywords{planet-star interactions; planets and satellites: dynamical evolution and stability}

\section{Introduction}

The search for potentially habitable planets around other stars has advanced significantly in recent years. As the precision
of radial velocity searches has improved, the detection limits have reached into the regime where planets are thought to be
too small to be genuine gaseous planets.
 Furthermore, estimates of the incompleteness of such searches 
suggest that planetary systems composed of low mass planets may be significantly more common than those with Jovian class
planets in short period orbits (Howard et al. 2010; Mayor et al. 2011). The launch of the Kepler satellite has
also unearthed a substantial catalogue of planetary candidates with radii indicative of small planets 
(Borucki et al. 2011; Batalha et al. 2013), including several candidates at sufficiently long periods to potentially qualify as habitable 
(Borucki et al. 2013; Barclay et al. 2013).

The origins of such plentiful systems of lower mass planets is still somewhat uncertain. There are a significant number of
planets in the Neptune-size category, which were predicted to be rare in population synthesis models based on the concept
of planetary migration (Ida \& Lin 2008; Mordasini et al. 2009). Furthermore, the period ratios in the multiple planet
systems (Lissauer et al. 2011; Fabrycky et al. 2012) are broadly distributed, contrary to predictions of migration 
models (Alibert et al. 2006; Terquem \& Papaloizou 2007; Raymond et al. 2008; Mordasini et al. 2009; Ida \& Lin 2010),
which anticipated a preference for commensurabilities. This has led to an alternative proposal that such systems assemble
in situ (Hansen \& Murray 2012; 2013; Chiang \& Laughlin 2013), although adjustments to the migration scenario have also
been proposed (Rein 2012; Goldreich \& Schlichting 2014). 

The bulk of the attention thus far has been focussed on solar-type stars, both for the obvious anthropocentric reasons and
because the observational samples are dominated by such stars. Lower mass stars, M-dwarfs, are disfavoured by virtue of
their lower luminosities, high activity levels,  and redder spectra (which reduce radial velocity accuracy). On the other hand, M-dwarfs are smaller
and therefore are attractive transit survey targets, as smaller planets obscure a larger fraction of the star than of a G-dwarf
during transit. Furthermore, the lower stellar luminosity also implies that the habitable zone is closer to the star and so
easier to study via transits because of the shorter orbital period and greater transit depths. As a result,
 several groups have recently begun 
 to quantify the sample of transiting systems that orbit M-dwarfs in the Kepler data (Muirhead et al. 2012; Mann 
et al. 2012; Dressing \& Charbonneau 2013).
Therefore, in this paper we explore the predictions of an in situ assembly model for M-dwarf planetary systems, adapting the
model of Hansen \& Murray (2013)--hereafter HM13--to the environs of a 0.5~$M_{\odot}$ star. We compare these to the sample of transiting
planet candidates defined by Dressing \& Charbonneau (2013)--hereafter DC13. In \S~\ref{Observe} we will quantify the observational sample
and use this to estimate the initial conditions necessary for our theoretical model. In \S~\ref{Model} we then describe the
construction of our theoretical model for in situ assembly and its comparison for the data.

\section{Observations}
\label{Observe}

Howard et al. (2012) noted an increasing incidence of short period, low mass planets as the effective temperature
of the Kepler host stars decreased. This suggests that the environs of M-dwarfs are an encouraging place to search
for planets. However, the stellar parameters for the coolest stars in the Kepler Input Catalogue are somewhat 
uncertain because the Kepler stellar characterisation was optimised for solar-like stars. Several groups have
recently attempt to remedy this by a variety of means. Muirhead et al. (2012) and Mann et al. (2012) have 
performed spectroscopic characterisations which reveal, amongst other things, that the brightest of the cool
Kepler stars are primarily giants, and that the dwarfs dominate the fainter part of the cool star luminosity function.
DC13 have performed a comprehensive comparison of stellar models with the available photometry to properly delineate
the cool star planet sample, and we will henceforth use their catalogue of planets and planet properties as our
comparison set. 

The resulting planetary catalogue is shown in Figure~\ref{PRcomp}, showing candidate period versus radius.
Solid points indicate planets in systems with multiple transiting planets, while the crosses indicate systems
 in which only a single planet transits. We see that the catalogue contains four planets with radii $> 3.5 R_{\oplus}$,
which appear to represent an extension of the hot Jupiter sample to low planet masses. Furthermore, we see that planets
with $P<$2.2~days are also overwhelmingly single (13/1 for $R< 3.5 R_{\oplus}$), in contrast to planets found at larger
radii. Anticipating the result that our model will lead primarily to multiple planet systems, composed of primarily
rocky planets, we will focus our attention on the sample enclosed by the dotted line in Figure~\ref{PRcomp}, namely
$P > 2.2$~days and $R<3.5 R_{\oplus}$.

\subsection{Protoplanetary Disk Parameters}

Our working hypothesis is that the final planetary system is a consequence of the gravitational assembly from an initial
disk of planetary embryos. Such a disk could arise from genuine in situ sedimentation, by the inspiral of smaller particles
by aerodynamic drag and subsequent growth, or by embryo inspiral due to protoplanetary nebula torques.
Our model is restricted to the final stage of assembly, which we assume 
is dominated by the gravitational interactions between the embryos.
In the case of planetary systems around G stars, the characteristic benchmark disk is the minimum
mass solar nebula (Weidenschilling 1977), although the currently observed systems often require substantially more mass than is
found in our own solar system (Hansen \& Murray 2012, HM13; Chiang \& Laughlin 2013).
In the case of M dwarfs, we need guidance from a different set of observations. To that end, we repeat the analysis of
Chiang \& Laughlin, but restricted to the cool star sample of DC13, in order to construct an equivalent initial density profile using the Kepler data. 

In order to derive a surface density profile $\Sigma = M_p/2 \pi a_p \Delta a_p$, we need to estimate $M_p$ from
the planetary radius and the relative proportion of $\Delta a_p/a_p$. Chiang \& Laughlin used the mass-radius relation from 
Lissauer et al. (2011), $M_p = (R_p/R_{\oplus})^{2.06}$. However, this is based on an interpolation between the terrestrial
planets and the ice giants of the solar system. The bulk of the planets in Figure~\ref{PRcomp} have $R<3 R_{\oplus}$, so
we expect these to be predominantly rock planets.
Furthermore, observations suggest that many low-mass planets have Hydrogen atmospheres large enough to significantly inflate
the radius while not increasing the mass substantially. Thus, we instead adopt a simpler estimate that all the planets in this
sample have mass $\sim 1 M_{\oplus}$, regardless of the observed radius.
 We also 
estimate $\Delta a/a \sim 0.4$. This latter number
assumes an average planetary mass of $1 M_{\oplus}$ around an average stellar mass of $0.5 M_{\odot}$, and that
the spacing, $\Delta$, between planets $\sim 50$, when measured in terms of mutual Hill radii. This is a conservative estimate,
as $\Delta \sim 50$ represents the most widely spaced pairs produced by our assembly simulations, but can partially
compensate for any underestimate in the planetary mass.
 Estimating $\Sigma$ in this fashion,
we find the profile shown in Figure~\ref{Sigplot}, and fit it with 
 a disk surface density profile
\begin{equation}
\Sigma (a) = 11 {\rm g cm^{-2}} \left( \frac{a}{\rm 1 AU} \right)^{-2}
\end{equation}
which yields a mass of 6~$M_{\oplus}$ from 0.05 to 0.5~AU. 

\section{Model}
\label{Model}

As in HM13, we begin our integrations at the end of the oligarchic assembly phase, using the 
equations of Kokubo \& Ida (1998) to specify the distribution of the initial planetary embryos in mass and semi-major axis.
For a density profile that drops off as the square of the distance, this results in a fixed initial embryo mass $\sim 0.26 M_{\oplus}$
and a total initial population $\sim 27$ (we allow for a small random fluctuation in the location of the inner embryo, which
in turn results in a small variation in the embryo mass across simulations in order to maintain the 6 $M_{\oplus}$ total disk
mass). We adopt an inner edge for the disk $\sim 0.05$AU, so that some small scatter in the final locations of observed planets
will produce a sample with $a>0.03$~AU (the inner edge of our observed sample). Initial orbits are assumed to be circular and
the dispersion in inclinations is $\sim 1^{\circ}$.
 The time step is now reduced to 6 hours because of the small inner edge. The central star is taken to be $0.5 M_{\odot}$
and the stellar radius to be $0.5 R_{\odot}$, in order to correspond to the median host mass in the DC13 sample. 
We also assume that there are no giant planets at larger distances, so that the rocky planets assemble only under their
mutual gravitational interaction.

We note that the chosen inner edge is larger than the dust sublimation radius calculated by Swift et al. (2012), who argued
against in situ formation based on the fact that a candidate planet in the Kepler-32 system is found interior to this radius.
However, the bulk of the planets in the multiple planetary systems sample lie well exterior to this boundary, suggesting that
Kepler-32f may be an outlier. The fact that it is also smaller than the other planets in the system argues in favour of a model
in which it was scattered inwards after building up to planetary size, as has been suggested for Mercury in our solar system
(e.g. Hansen 2009).

\subsection{Integrations}
\label{Scat}

We integrate 50 realisations of this model using the Mercury integrator (Chambers 1999). The simulations are run for
$10^7$~years and then we examine the ensemble properties, several of which are tabulated in Table~\ref{OutTab}. Figure~\ref{MultiM} shows the distribution of the number of
surviving planets with semi-major axis $<0.5$~AU. We see that most systems comprise between 4 and 6 surviving planets inside 0.5~AU, although the 
largest number goes as high as 10. Figure~\ref{StatComp} compares the result of these simulations with 50 realisations of the model of HM13 -- the
equivalent for solar-type stars. We see that the M-dwarf planetary systems have separations similar to the solar-type case (when measured in
terms of mutual gravitational influence), but are somewhat more circular and coplanar. This also indicates that the system configurations are dynamically
stable at this point, because there is not sufficient angular momentum deficit in the systems (Laskar 1997) to facilitate orbit crossing.

We use the results of these simulations to generate input distributions for a Monte Carlo population synthesis model.
 Figure~\ref{MbinM} shows the mass distribution of the surviving planets. There is
little trend of planet mass with semi-major axis, and so we model this distribution as
\begin{equation}
p_m(x) \sim x^2 {\rm exp}(- \frac{1}{2} (x/0.65 {\rm M_{\oplus}})^2)
\end{equation}
as shown in the figure. We have not attempted to match all the features of the observed histogram as some of the structure
is the result of the coarseness of our mass resolution (recall our embryo mass $\sim 0.26 M_{\oplus}$).

The spacing of the final planets is similar to those of the higher mass systems (HM13), when expressed in units of the
mutual Hill radius, i.e. $\Delta = 2 (a_2-a_1)/(a_2+a_1)/((M_1+M_2)/3 M_*)^{1/3}$. Figure~\ref{DeltaM} shows the
resulting distribution of planet pairs. We see that the distribution peaks between $\Delta \sim $20--25, and spans
the range $12<\Delta<45$.
For the purposes of the Monte Carlo model, however, the distribution of orbital separations
 $\delta = 2 (a_2-a_1)/(a_2+a_1)$ is used, namely 
\begin{equation}
p_{\delta}(x) \sim {\rm exp}\left(-\frac{1}{2} \left( \frac{x-0.4}{0.2} \right)^2 \right).
\end{equation}

The inclination dispersion, relative to the original orbital plane, is shown in Figure~\ref{ibin}, and is well characterised by
\begin{equation}
p_i(x) \sim x \, {\rm exp}(-x/1.3^{\circ}).
\end{equation}
The probability of transit is determined not only by the distribution of inclinations, but also the distribution of
the lines of nodes relative to the observer. Figure~\ref{DeloM} shows the distribution of $\Delta \Omega$, the difference
in the position of the line of nodes for neighbouring pairs in our simulations, at 10~Myr. We characterise this as a
flat distribution in $\Delta \Omega$, plus enhancements by a factor of 1.7, for $0<\Delta \Omega<10^{\circ}$, 
$99^{\circ}< \Delta \Omega < 109^{\circ}$, and $173^{\circ} < \Delta \Omega < 180^{\circ}$.

The model for the distribution of the mutual offset in the nodes of neighbouring planets is the same as used in HM13.
Eccentricities are small, and well characterised by $p_e(x) \propto x\, {\rm exp}(-x/0.03)$.

\subsection{Monte Carlo Model}

As in HM13, we use the distributions $p_m$, $p_e$, $p_i$ and $p_{\delta}$, fit to the simulations, to construct a Monte Carlo model for
the underlying planetary population, in order to provide a population synthesis model that can
be observed in the same manner as a transiting planet search. We draw from these distributions the separations, masses, inclinations,
eccentricities and nodal alignments of model planetary systems. We remove any which violate our criteria that $12<\Delta<40$, and then
determine what fraction of the planets will be observed as transits from observers at various orientations. We assume the stellar radius
is $0.5 R_{\odot}$ and also adopt a parameter $R'$ that specifies the degree of planetary radius enhancement, in order to account for
the possibility that the radii are inflated by Hydrogen atmospheres.
The detection efficiency for the DC13 sample can be modelled in a similar fashion as Youdin (2011) did for the
sun-like sample, using the numbers in Figure~15 of DC13. We find that the detection efficiency for planets
is not strongly dependant on planetary period, and that the recovery fraction $f_{recover}$ can be approximated as 
\begin{equation}
f_{recover} = 1 - {\rm exp}(-0.7 (R/{\rm R_{\oplus}})^2).
\end{equation}
We use this function to estimate the probability that a transiting planet in our Monte Carlo sample is
actually recorded as a tranet, a term we adopt, following Tremaine \& Dong (2012), to avoid confusion
between the sample observed in a transit survey (tranets) and the underlying population (planets). 

The first measure of comparison is the ratio of tranet systems of different multiplicity.
Figure~\ref{MultiComp} shows the results from the Monte Carlo model, compared to several observational
determinations. The open circles show the ratio determined from the full DC13 catalogue, and the
crosses indicate the ratios in the catalogue of Muirhead et al. (2012). The filled circles show
the ratios calculated using only the sample contained in the dotted line in Figure~\ref{PRcomp}. The
exclusion of 17 singles (although one is regained since one double loses a member too) increases the
2:1 ratio slightly, but not sufficiently to change any qualitative conclusions. As for the G~star
sample (see HM13), the ratio of higher multiples matches the model (although the statement is weaker
in this case because of the smaller sample size) but we see again an excess of single tranet systems
which suggests a substantial population of systems whose multiplicities must be lower than in our model.
Models with a modest enhancement in radius $R'>1.2$ are consistent with the observed multiplicities and
we adopt $R'=1.4$ as our default model. We also show a comparison with a model in which $R'=1.4$ only
for $a>0.07$AU and is set to $R'=1$ interior to that. This is an approximation to the idea that the
closest planets may lose some or all of their Hydrogen atmospheres due to evaporation, driven
by the stellar flux (Owen \& Jackson 2012; Lopez, Miller \& Fortney 2012; Owen \& Wu 2013; Lammer et al. 2013).

The histogram in Figure~\ref{PratM} shows the distribution of period ratios for neighbouring tranets
from our Monte Carlo model. The points show the same quantity for the multiple tranets in the DC13
catalogue. The excellent agreement further supports the basic notion that the spacing of the planets
in the multiple tranet systems observed by Kepler are well represented by an in situ assembly model.
We also denote the location of the 2:1 and 3:2 commensurabilities, in order to demonstrate that the
M~dwarf sample shows no significant preference for such period ratios. There is not even the slight excess
observed in the G-dwarf sample (Lissauer et al. 2011; Fabrycky et al. 2012),  although the number of observed
systems is also smaller.

The distribution of periods themselves is also well produced by the models, as seen in Figure~\ref{Pdis}.
The solid points show the periods of tranets in multiple systems, while the open circles indicate the
distribution of single tranets. The dashed histogram shows the prediction of a model with $R'=1.4$. This
fits most of the data well, but overpredicts the number of planets with periods $< 6$~days. 
 The solid histogram shows the distribution drawn from the model in
which we use $R'=1.4$ for $a>0.07$AU only, and $R'=1$ interior to that. We see
that the general shape is now well matched.
 An alternative way to match the innermost bins would be to have a slight distribution
of inner edges to the original disk, as our present model is rather naive with a uniform inner edge.

Figure~\ref{DelM} shows the comparison between the models and the observations when the spacings are
normalised by their mutual Hill radii. The histogram shows the model distribution of $\Delta$, for $R'=1.4$. To get
an observational distribution, we need to convert the observed radii to masses. a common procedure in
the literature is to use the
relationship of Lissauer et al. (2011). We designate the $\Delta$ calculated with this relationship as $\Delta_L$. Much as for the G~star sample, the
observed distribution of $\Delta_L$ shows a similar shape as that from the simulations, but shifted
to lower numerical values. This once again suggests that the masses estimated using Lissauer's relation
are overestimated. In Figure~\ref{DelM} we have multiplied $\Delta_L$ by a factor of 1.5 to match the 
simulated distribution, which corresponds to the assumption that the Lissauer relation leads
to a mass overestimate by a factor of 3. For a planet of radius
$2 R_{\oplus}$, this suggests that masses should be more like $1.4 M_{\oplus}$ rather than $4.2 M_{\oplus}$.
 Regardless of the difference in normalisation, we can see that the
shapes of the observed and theoretical distributions are similar, with a broad peak and a long tail to
higher values.
This again implies that some observed neighbouring tranet pairs have inferred $\Delta$ well above the
expected maximum from the simulations, suggesting that additional planets await detection in the gaps. 
The four systems in the DC13 table that satisfy this criterion are 936.01\& 936.02, 2719.01 \& 2719.02,
2650.01 \& 2650.02 and, finally, 1422.01 \& 1422.02. None of these pairs are particularly surprising
in this regard, since all of them show period ratios in excess of three.

Finally, the mutual inclinations of neighbouring planets can also prove to be an interesting constraint
on the models (e.g. Fabrycky et al. 2012), and is well matched by the in situ assembly model for solar-type
stars (HM13). Unfortunately, the DC13 sample is sufficiently small that the shape of the distribution is
not well constrained observationally, especially since the impact parameters of individual tranets are
only reported to a single significant digit.

\subsection{Habitability and Water Delivery}

A particularly exciting aspect of planetary systems around M dwarfs is the possibilities they offer
for studying the habitability of planets through transit observations, by virtue of the proximity
of the habitable zone (HZ) to the host star. Figure~\ref{Teqp} shows the
sample of known M dwarf planets and candidates as a function of distance and host star effective temperature,
along with estimated locations of the HZ using the recent estimate of Koppurapu et al. (2013).

However, the habitability of such planets is far from assured, as the proximity to the host star also introduces
additional effects, such as the tidal locking of the planetary spin, and the consequent possibility of dramatic 
atmospheric effects (Kasting et al. 1993, Joshi et al. 1997). An additional problem is that
such planets may be quite dry. Raymond et al. (2007) and Lissauer (2007) have argued that it will be harder for planets in M-dwarf HZ
 to accrete volatiles, including water, because they are well separated from the ice line and would have
difficulty accreting in high velocity impacts. 
It has long been appreciated that, in our Solar System, the condensation of water into solid material suitable
for planetary accretion requires lower temperatures than expected in the region where Earth accreted. This has led
to a long debate regarding the origins and timing of the delivery of water to the Earth (Morbidelli et al. 2000; Raymond
et al. 2007; Izidoro et al. 2013). Therefore, 
 we have performed a set of simulations with a more extended disk and a population of test particles to
trace the dispersal of water-bearing small bodies across the final planetary system. 

Our model uses the same disk profile as in \S~\ref{Scat}, but now extended out to a distance of 1~AU. If we scale
from the solar system ice line at $\sim 2.5$~AU, to a 0.5$M_{\odot}$, $T_{eff}=3800$K M-dwarf representative
of the DC13,
 we estimate that water ice will begin to form at distances $\sim (3800/5800)^2 \times 2.5/2 = 0.54$~AU,
 and so we include 100 test particles in each simulation, distributed uniformly in semi-major axis from 0.54~AU and
1~AU. This weights the larger separations more than the underlying oligarchic mass distribution, and represents a crude
representation of increasing condensation at larger distances and cooler temperatures.

For our 0.5$M_{\odot}$ star, a moderately conservative estimate for the  HZ lies between 0.23--0.44~AU (using the Koppurapu models
for a runaway greenhouse at the inner edge and an early Mars for the outer edge). 
Our simulations produce either one or two
surviving planets in this range, at an age of 10~Myr. Figure~\ref{Water} shows the locations of the surviving bodies for
five realisations of our model. For the planets in or near the HZ, we also include two labels showing what fraction
of the original water-bearing planetesimals were accreted by the planets. The upper label indicates the total fraction
of the test particles accreted by the surviving body, including those accreted by embryos that were later accreted. The
lower label indicates only that fraction of the total test particle inventory that was accreted after the last embryo
was accreted by this planet. The difference between the two numbers represents the range of possibilities, depending on
whether volatiles are lost during giant impacts (lower number) or not.
 We see that there is substantial scatter in the amount of water accreted, ranging from 0 to
23\% of the original reservoir. We also note that many systems contain a `gatekeeper' body located just outside the HZ, which accretes a
substantial fraction of the water.

Lissauer (2007) also raised the possibility that even late accretion from water-rich bodies might not be very efficient because
of the high velocities of encounter. In our case, the presence of gatekeeper bodies will scatter bodies on orbits that
penetrate the inner planetary system. This will potentially serve to limit the flux of such bodies but may also make their
evolution more diffusive and therefore reduce the velocities of encounter for those that do penetrate the HZ.
To estimate this effect, Figure~\ref{Vesc} shows the final encounter velocities, normalised to the escape velocity
from the accreting rocky planet, for those test particles accreted after the
last major impact, for planets with final semi-major axis between 0.17 and 0.38~AU and summed over 10 simulations. We further divide the
habitable zone into the inner (IHZ) and outer (OHZ) parts, with the transition at 0.27~AU. In both cases, approximately 50\%
of the accreted material is acquired through encounters with less than twice the planetary escape speed -- the criterion of
Melosh \& Vickery (1989) for substantial atmosphere erosion. This suggests that the build-up of volatiles on these planets
can be very stochastic, with approximately equal chances of gaining or losing from the planetary inventory with each encounter.
The planets in the IHZ might have a slight advantage in this regard, as the tail to lower velocities is larger, presumably reflecting 
the more diffusive orbital evolution of planetesimals that manage to penetrate that far inwards.

These numbers demonstrate that a sufficient water inventory is potentially possible for many of these habitable planets. 
The water inventory of the Earth is estimated at a mass $\sim 2 \times 10^{-3} M_{\oplus}$ (Drake \& Campins 2006; Marty 2012), with
about 10\% contained in the oceans. If we assume that 50\% of the
late accreted water in these simulations is retained, then IHZ planets capture and retain $\sim 3\%$ of the original reservoir,
while OHZ planets retain $\sim 2\%$ on average. Estimates of the condensible fraction of material in a protoplanetary disk suggest
that, beyond the ice line, the mass inventory approximately doubles (e.g. Hughes \& Armitage 2012).  If the same holds for
the model discussed here, this implies 
 as much as $2 M_{\oplus}$ of water is available in planetesimals.
 This implies that the HZ planets around white dwarfs can accrete $\sim 0.05 M_{\oplus}$ of water, an order of magnitude
more than estimated for the water inventory of Earth. The actual reservoir of water will depend on the physical processes
that produce the initial conditions for final assembly, but the results here show 
 that planets in M~dwarf habitable zones can be large enough, and accrete sufficient
material at late times to potentially also meet the water delivery requirement for habitability. 

\section{Discussion}

M~dwarfs have a history of confounding pessimistic theory predictions regarding their planetary inventories.
Laughlin et al. (2004) predicted that the mass and timescale requirements for giant planet formation via core
accretion would make giant planets relatively rare around M~dwarfs. Although this is partially born out by the
relative rarity of giant planets with short orbital periods around M dwarfs (Johnson et al. 2007), some have
been discovered (Rivera et al. 2010; Haghighipour et al. 2010; Johnson et al. 2010) and 
 microlensing surveys suggest that many
M~dwarfs may indeed possess giant planets at larger distances (Gould et al. 2010). Similarly, Raymond et al. (2007)
predicted that protoplanetary disks which scaled linearly with stellar mass would produce terrestrial planet
systems with planets too small ($< 0.3 M_{\oplus}$) to be detected by Kepler in significant quantities, although
Montgomery \& Laughlin (2009) demonstrated that more massive disks could indeed produce planetary systems of
observable size via in situ accretion. Furthermore, planetary systems may also be constructed via a number of
other evolutionary pathways beyond simple in situ accretion (Raymond et al. 2008, Ogihara \& Ida 2009), which 
could potentially alleviate the mass constraints.

The Kepler mission has now revealed that planetary systems around cool stars follow a similar trend to those
around hotter stars, in the sense that compact systems of low mass planets appear to be more numerous than
those containing gas giants, at least at short orbital periods.
 Our goal in this paper has been to demonstrate that these low mass planetary systems are
indeed broadly consistent with a population of planets that assembled in situ, suggesting a commonality with the
planetary systems observed around sun-like stars (Hansen \& Murray 2012, HM13; Chiang \& Laughlin 2013). This
is somewhat at odds with the conclusion reached by Swift et al. (2013), who favour inward migration of planets. 
However, that study relied on a detailed
study of a single system (Kepler-32), allied to blithe generalities about the rest of the Kepler sample. The
Kepler-32 system is one of the few observed systems that actually corresponds to the commensurable chain expected
from migration models, and therefore cannot really be used as a representative of the sample as a whole 
 (consider the distribution
of  period ratios shown in Figure~\ref{PratM}). Furthermore, the particular arrangement of the Kepler-32 system may 
simply be a statistical fluctuation. Figure~\ref{Zeta} shows the distribution of the four Kepler-32 period ratios when characterised
by the statistic $\zeta_i$ introduced by Lissauer et al. (2011), to characterise the proximity to a resonance of order $i$.
The open circles indicate the four Kepler-32 pairs, while the filled circles show the rest of the pairs in the DC13 sample.
We see that the Kepler sample as a whole is spread out over the full observed range of $\zeta$, and could easily represent a generic
sampling of the fuller distribution.
 Also shown is the distribution of $\zeta_1$ expected from our
Monte-Carlo model, which reproduces the broad distribution seen in the observations. In short, in situ assembly reproduces
the observed broad distribution of period ratios quite well, even with the commensurabilities of the Kepler-32 system
included in the sample. Migration may still be required to explain systems with giant planets (Haghighipour 2013) however.

The amount of mass used in our rocky nebula model is $6 M_{\oplus}$ between 0.05~AU and 0.5~AU. If we estimate the same
amount of mass in this range using the surface density profile assumed in HM13, we get 12.4~$M_{\oplus}$, around a
star of twice the size. Thus, our two models nicely match the approximate linear scaling of disk mass with respect to
stellar mass inferred by Andrews et al. (2013) from observations of infrared excesses around stars in Taurus. 
Furthermore, this amount of rocks requires only $\sim 10^{-3} M_{\odot}$ of gas for a solar metallicity protoplanetary disk,
and is thus quite possible around a young M-star.

Our model also has implications for the detectability of habitable planets around M~dwarfs. We have shown in Figure~\ref{Water}
that the
disk mass appropriate to explaining the observed distribution of tranets usually produces one, and often two, planets in the
nominal HZ. We can generalise this assertion by making use of the same semi-analytic estimate of planetary spacings
outlined in HM13. If we assert that a given surface density profile results in a chain of planets that are characterised by mass,
energy and angular momentum conservation, and the requirement that the final planets be spaced according to the $\Delta$ distribution
shown in Figure~\ref{DeltaM}, then we can relate the $\Delta$ between two neighbouring planets to the quantity $y=a_2/a_1$, where
$a_1$ and $a_2$ are the semi-major axes of the inner and outer members of the pair. Using the same $\Sigma \propto R^{-2}$ profile,
we derive
\begin{equation}
\Delta = 115 \frac{y-1}{y+1} \left( \frac{M_*}{0.5 \rm M_{\odot}} \frac{6 \rm M_{\oplus}}{M_{disk}} \frac{1}{\ln y} \right)^{1/3}
\end{equation}
where $M_{disk}$ is the total planetesimal disk mass, integrated from 0.05~AU to 0.5~AU. If the habitable zone extends from
0.23 to 0.44~AU, this corresponds to y=1.91, and $\Delta \sim 42$. A comparison with Figure~\ref{DeltaM} shows that this value is 
larger than almost all of the pairs in our simulations, and indicates that is  very difficult to form a system with this model without at least
one planet in the HZ.

We can extend this discussion by asking how large a disk do we need before it becomes likely to not have any planets in the
HZ? If we set $\Delta=25$, the peak of the distribution, and keep $y=1.9$, we can use this to derive a disk mass (between 0.05 and 0.5~AU) of $M_{disk}>28 M_{\oplus}$. 
A more useful comparison is if we convert this disk mass into the summed mass of the pair of planets that might straddle the
HZ. For our surface density profile, the mass fraction of the total 0.05--0.5~AU disk that is contained in the HZ
is 0.28, so that a pair of planets that straddles the habitable zone and has a combined mass $>7.8 M_{\oplus}$ may not have
any further planets between them.

Our discussion has thus far focussed on transiting systems, because of the larger sample size and prospects for eventual
atmospheric characterisation. However, several planetary systems have also been discovered around M~dwarfs, and these are
also well matched by in situ assembly models. Figure~\ref{Stat2} shows the comparison of the simulation results with
the observed planetary systems around
the stars GJ~581 and GJ~667C (Udry et al. 2007; Mayor et al. 2009; Vogt et al. 2010; Bonfils et al. 2011; Forveille et al. 2011; Delfosse et al. 2012; Vogt, Butler \& Haghighipour 2012; Anglada-Escude' et al. 2012, 2013). 
We compare the relative spacings ($S_s$) and mass concentration indices ($S_c$) shown in
Table~\ref{OutTab} with the mass and periods measured from radial velocities. We see that both the observed systems fall within the parameter 
range probed by the simulations. The star GJ~667C hosts several potentially habitable planets (Anglada-Escude' et al. 2013), as expected
from the above discussion. We can also adapt this rationale 
 to the controversy regarding the existence of habitable
planets in the GJ581~system (Vogt et al. 2010, 2012; Forveille et al. 2011). The existence
of GJ~581c and GJ~581d are not disputed, and they exhibit a period ratio of 6.8, corresponding to $y=3.59$. Their estimated
minimum combined mass is 11.2 $M_{\oplus}$, which yields $\Delta \sim 34$, for a host star mass $\sim 0.32 M_{\odot}$. The
issue is whether the claimed planet GJ581g exists between them. Comparing to Figure~\ref{DeltaM}, we see that $\Delta \sim 34$
is larger than most pairs, but does occur in a few cases. Thus, it is not impossible that there is a true gap between GJ~581c and GJ~581d,
although there should be another planet between them in the majority of planetary systems of this type. If GJ~581g did exist
with the nominal $3.1 M_{\oplus}$ mass and 36.6~day period, it would yield $\Delta \sim 23$ and $\Delta \sim 12$ with respect
to the inner and outer companions, which is also not ruled out on dynamical grounds, although it would be comparable to the 
most compact systems that emerged from the simulations. It is possible to locate a planet in this gap in a more 
 dynamically stable location, in terms of maximising the $\Delta$ values
with respect to both GJ~581c and GJ~581d. Planets with  periods $\sim 29$~days would have $\Delta \sim 20$ with respect to both 
neighbours,  assuming a $1 M_{\oplus}$ planet in the gap.

Although the in situ assembly model provides a good description of the distribution of period ratios and overall period
distribution of the observed multiple systems, we must also note that the model once again underpredicts the number of
single tranets observed, just as in the case of the solar-like hosts (HM13). Furthermore, Figure~\ref{MultiComp} shows that
the ratio of singles to doubles is similar  to the case of the solar hosts. However, unlike in the solar case, it does appear as though there
is some difference in the distribution of single tranets, in that there is a great overabundance of single tranet systems for
orbital periods $< 2.2$~days. Once again, this discrepancy suggests that our description is not complete, and that a comprehensive
model for the formation of these planetary systems will require additional elements to either reduce the multiplicity of some systems, or to increase
the spread in inclinations. Indeed, if we postulate a second process that produces only single planet systems, then, in order
to match the observed multiplicities, we require that
$\sim 58\%$ of observed tranet systems form via this alternative process. Our model would then be responsible for the remaining 42\%  systems, and 
for 56\% of all observed tranets, including all multiple tranet systems.

This has implications for the estimates of the frequency of habitable planets around M~dwarfs. The analysis of the frequency of 
tranet detections (DC13, Kopparapu 2013) indicate that such planetary systems around M~dwarfs
are quite common, and that the nearest planet in the HZ of an M~dwarf may be within only a few parsecs. DC13 estimate that 
25\% of early M-dwarfs host a planet with radius between 0.5-1.4 $\rm R_{\oplus}$. 
 If we assume that these planets result from our in situ
model plus the second, singles-only process in the proportions above, this implies that 10\% of such stars host
 high
multiplicity systems. 
Given that every planetary system formed by the above model produces at least one habitable zone planet, this suggests that
the incidence of habitable planets is $\sim 0.10$ per star, which is quite similar to the number obtained by DC13. It would
also increase if we include those systems with radius between 1.4--4 $\rm R_{\oplus}$, which have an equivalent occurence rate.
 Our results therefore only
strengthen the conclusion of DC13, in that we demonstrate that the bulk of the observed systems are well reproduced by an in situ
assembly model, and that the model suggests that a large fraction of planetary systems observed to have a transiting earth-mass planet at any
period should
also have at least one planet within the HZ.

\acknowledgements 
This research has made use of the NASA Exoplanet Archive, which is operated by the California Institute of Technology, under contract with the National Aeronautics and Space Administration under the Exoplanet Exploration Program, and has also benefitted from the considerable governmental and community investment of
resources and effort that produced the spectacular scientific legacy of the Kepler mission.

\newpage


\clearpage
\plotone{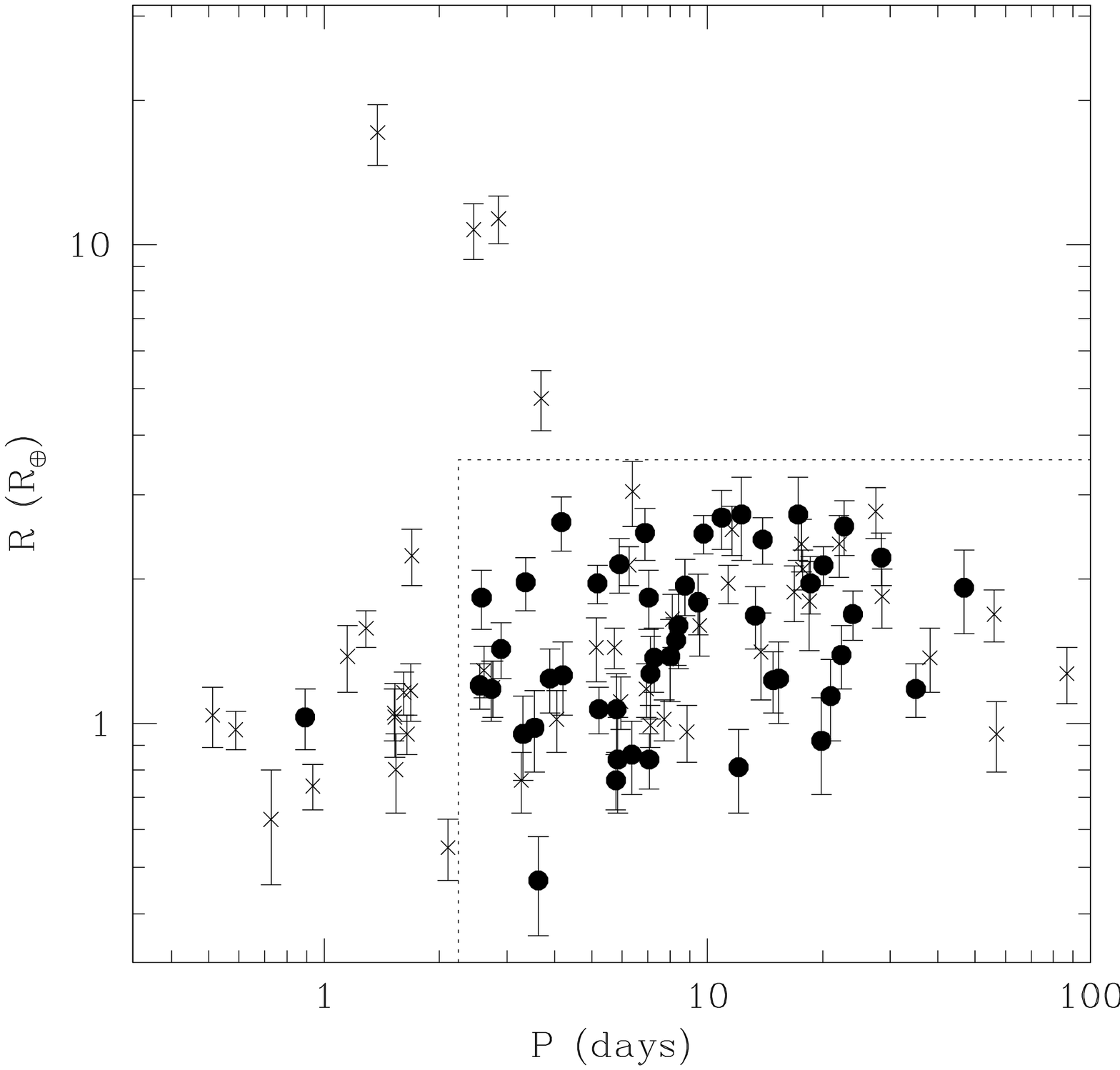}
\figcaption[PRcomp.ps]{The solid points show the planets from the DC13 catalogue that are found
in multiple systems. The crosses show the planets found in single systems. The dotted line 
encloses the sample used to calculate the multiplicity statistics used in the text to compare the model
to observations.
\label{PRcomp}}

\clearpage
\plotone{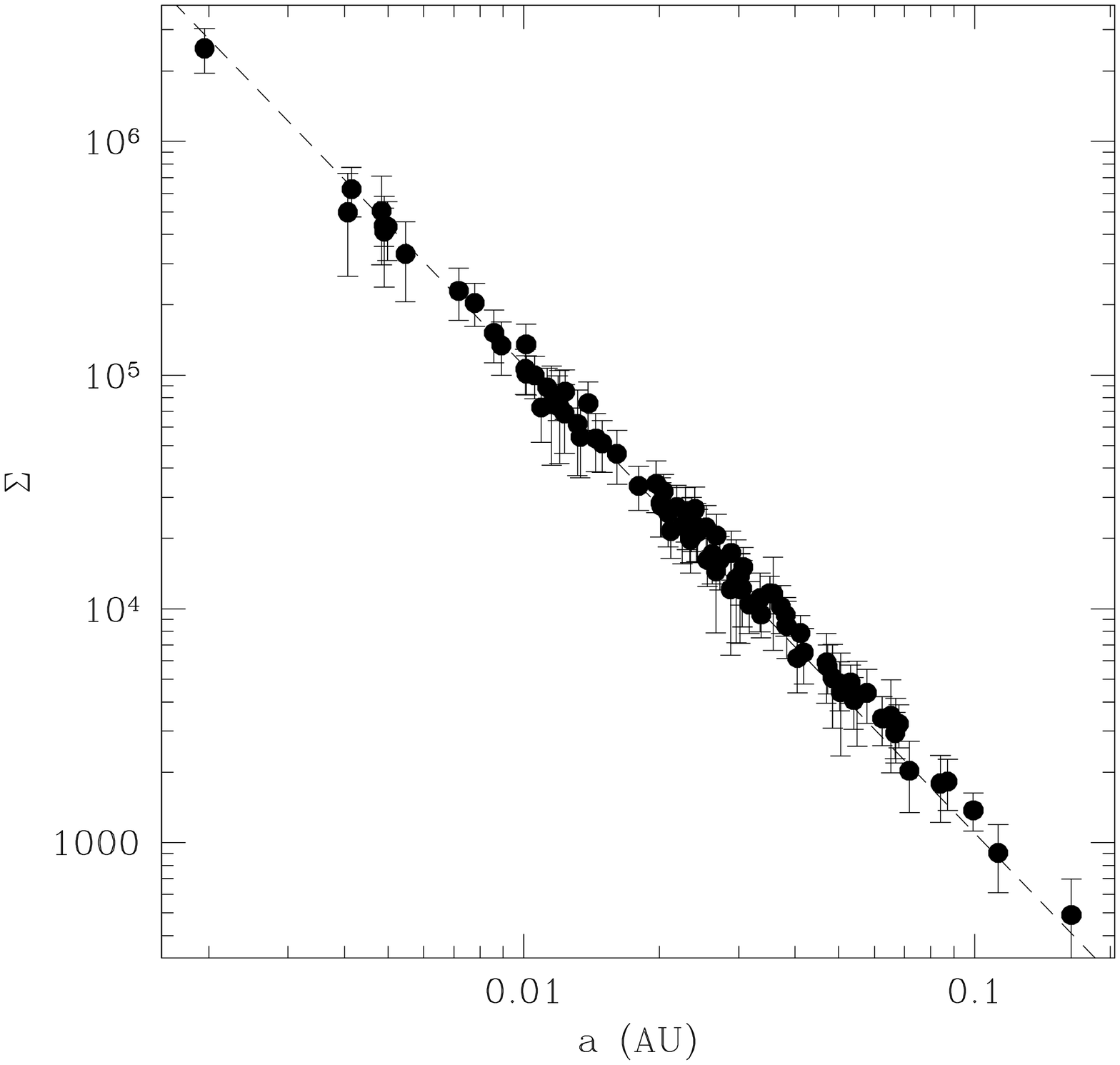}
\figcaption[Sig.ps]{The points represent the surface density $\Sigma$ calculated for each of
the planets in the DC13 catalogue, as described in the text.  The resulting
 $\Sigma \propto R^{-2}$ model is shown as a dashed line.
\label{Sigplot}}
\clearpage

\plotone{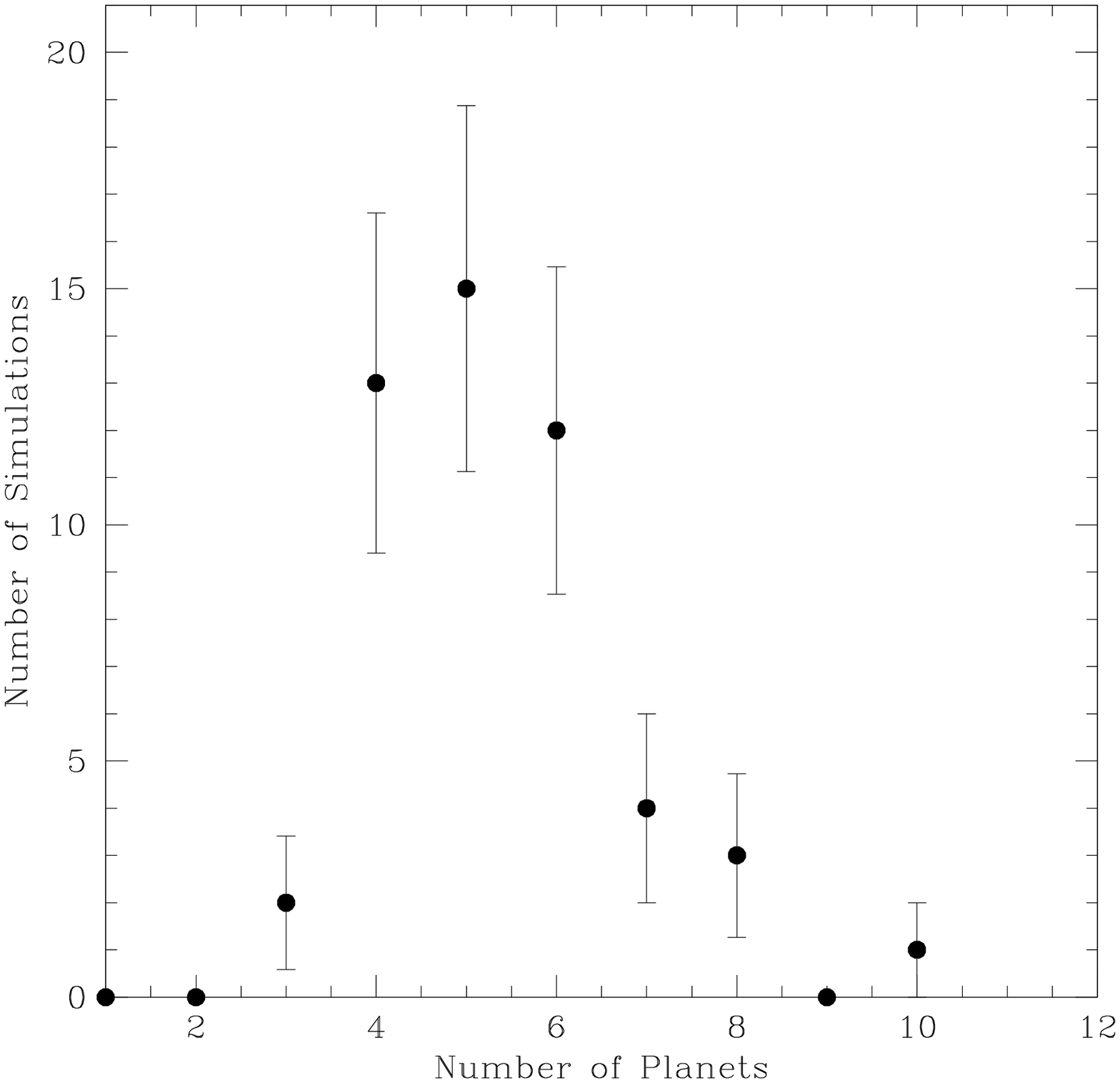}
\figcaption[MultiM.ps]{The points indicate the distribution of the final multiplicities of
the simulated systems. For each simulation we calculated the number of surviving planets
with semi-major axis $< 0.5$AU, after 10~Myr. Every system contains at least three surviving
planets, with most systems having between four and six.
\label{MultiM}}

\clearpage

\plotone{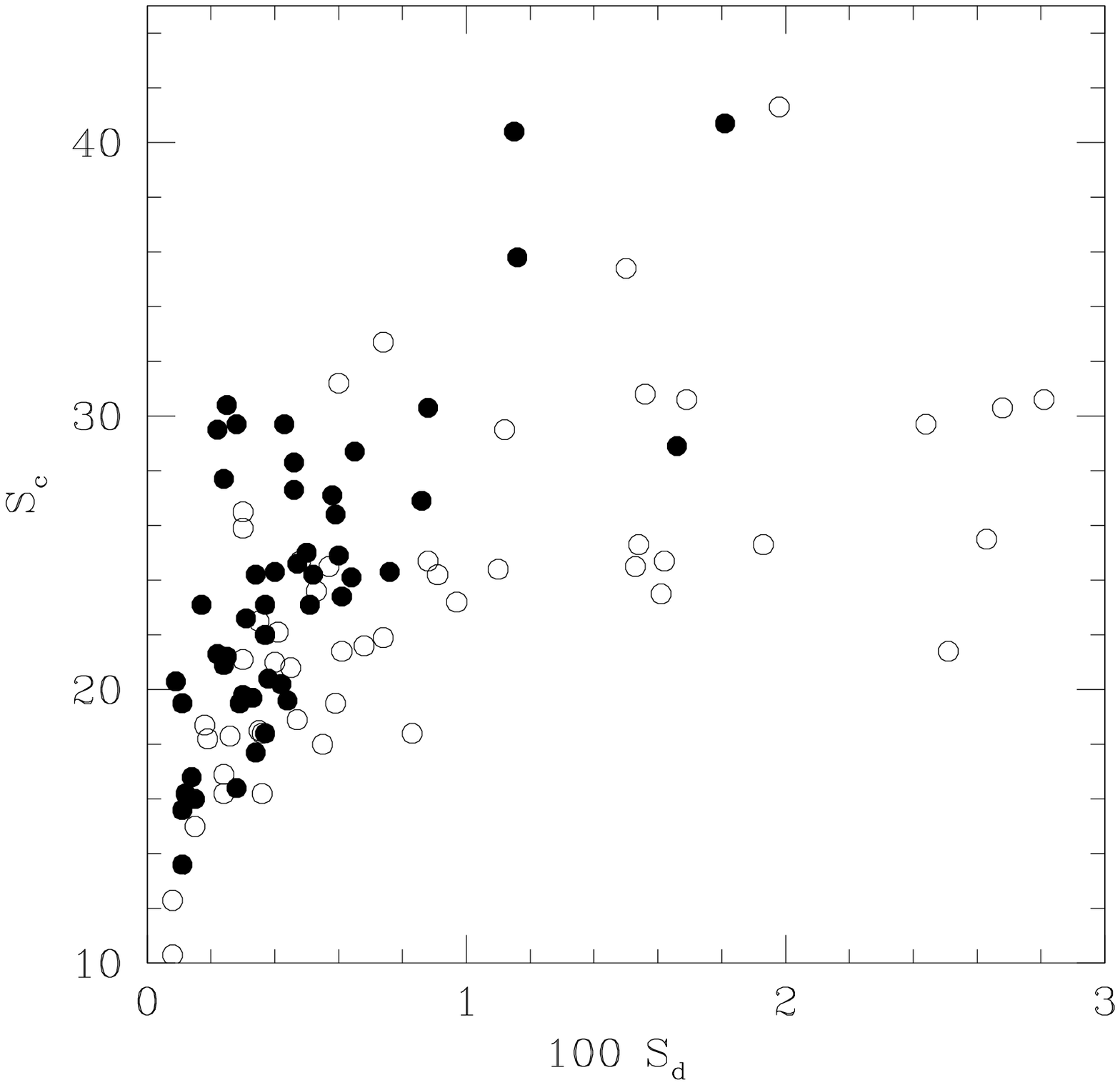}
\figcaption[StatComp.ps]{The open circles show the statistics $S_d$ (a measure of how circular/coplanar
a planetary system is) and $S_c$ (a measure of how closely packed a system is), for 50 realisations of
the planetary system model in HM13. The filled circles are the same measures but for the 50 realisations
of the model in this paper. We see that the M-star planetary systems are more circular and coplanar but
with similar spacing.
\label{StatComp}}

\clearpage

\plotone{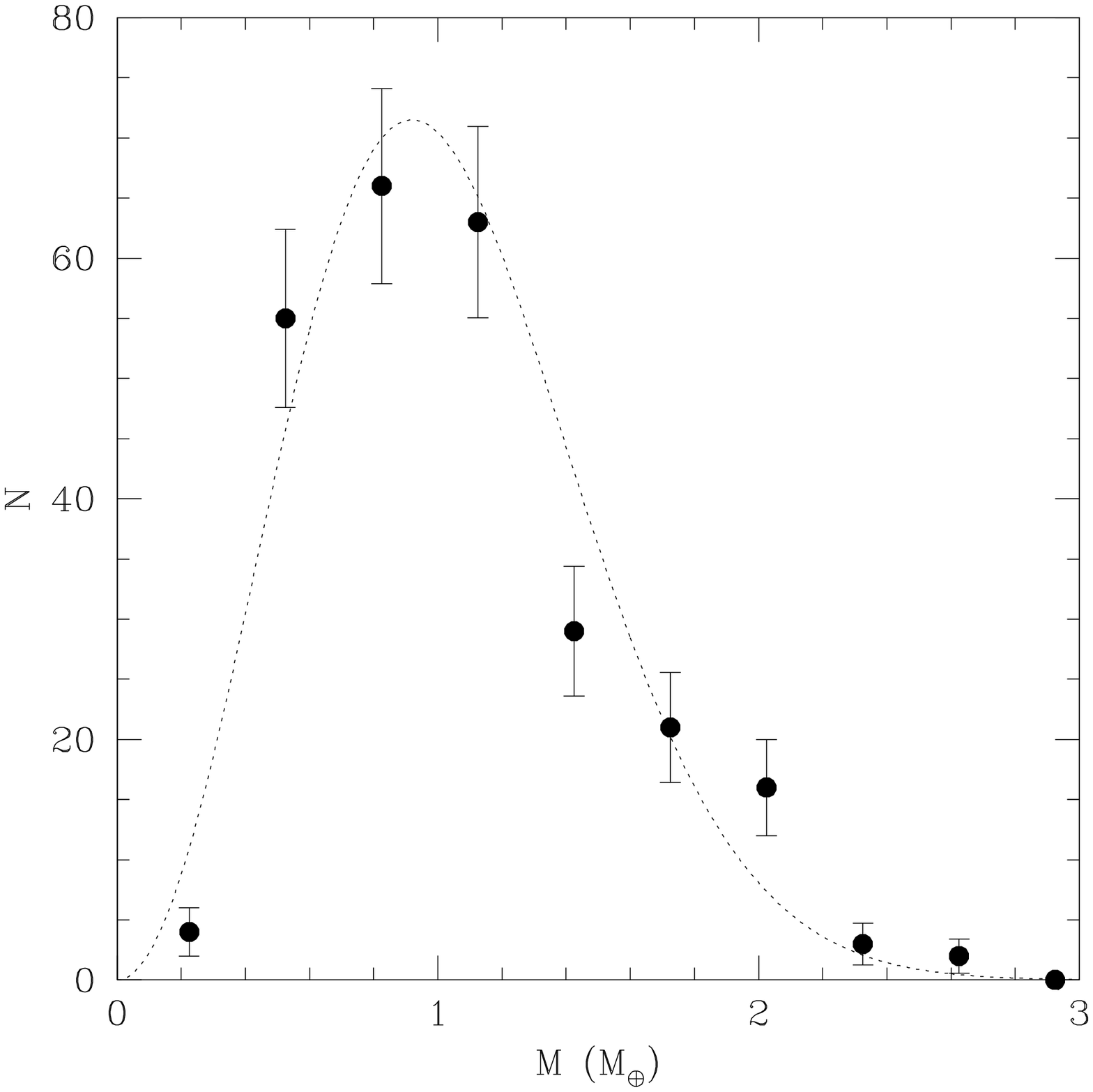}
\figcaption[Mbin.ps]{The points are obtained by binning up the surviving planets inside 1~AU over
all the simulated runs. There is no significant trend with semi-major axis. The dotted line is
a model of this distribution that is used to generate the Monte Carlo model. The shape of the
distribution is reminiscent of that inferred from the observed radius distribution (DC13, Morton \& Swift 2013).
\label{MbinM}}

\clearpage

\plotone{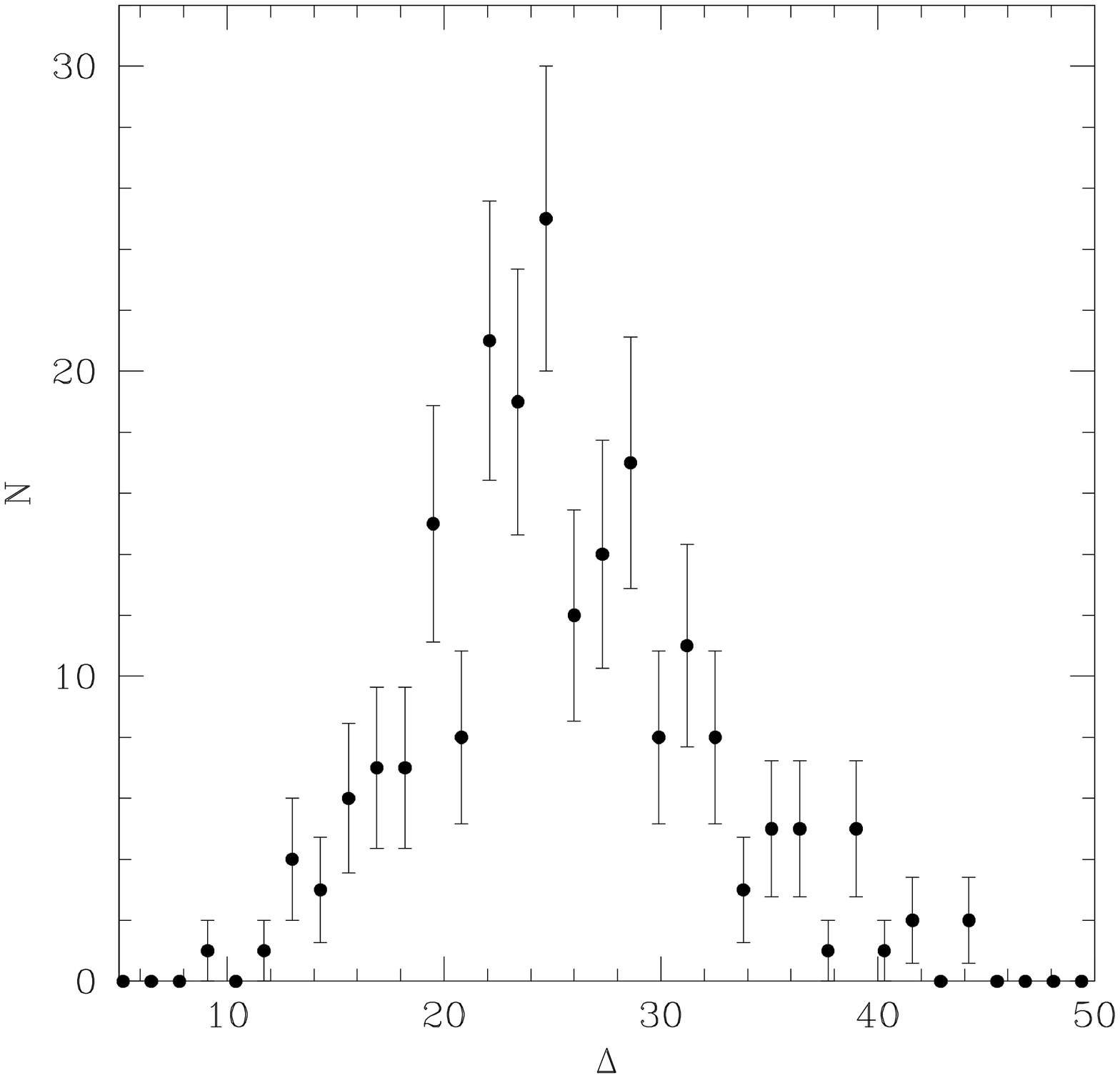}
\figcaption[DeltaM.ps]{The points show the distribution of $\Delta$ for planet pairs in the simulations
which survive to 10~Myr, and have $a<0.5$AU. The peak at $\Delta \sim 25$ is similar to the distribution
obtained for higher mass systems in HM13.
\label{DeltaM}}

\clearpage

\plotone{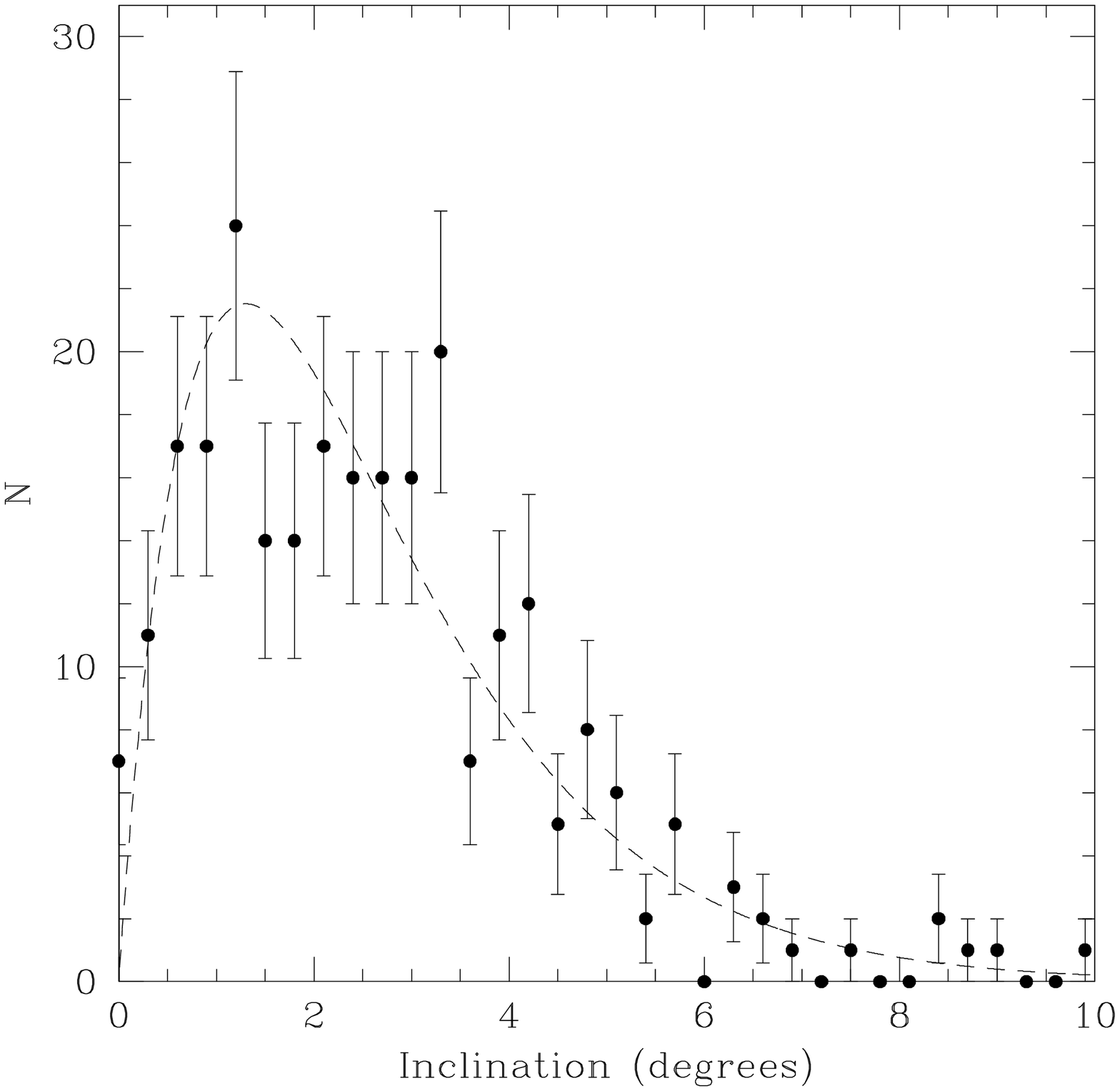}
\figcaption[ibinM.ps]{The points show the inclinations of surviving planets in the simulations, measured relative to the original
orbital plane. The dashed line is used to simulate this in the Monte Carlo model.
\label{ibin}}

\clearpage

\plotone{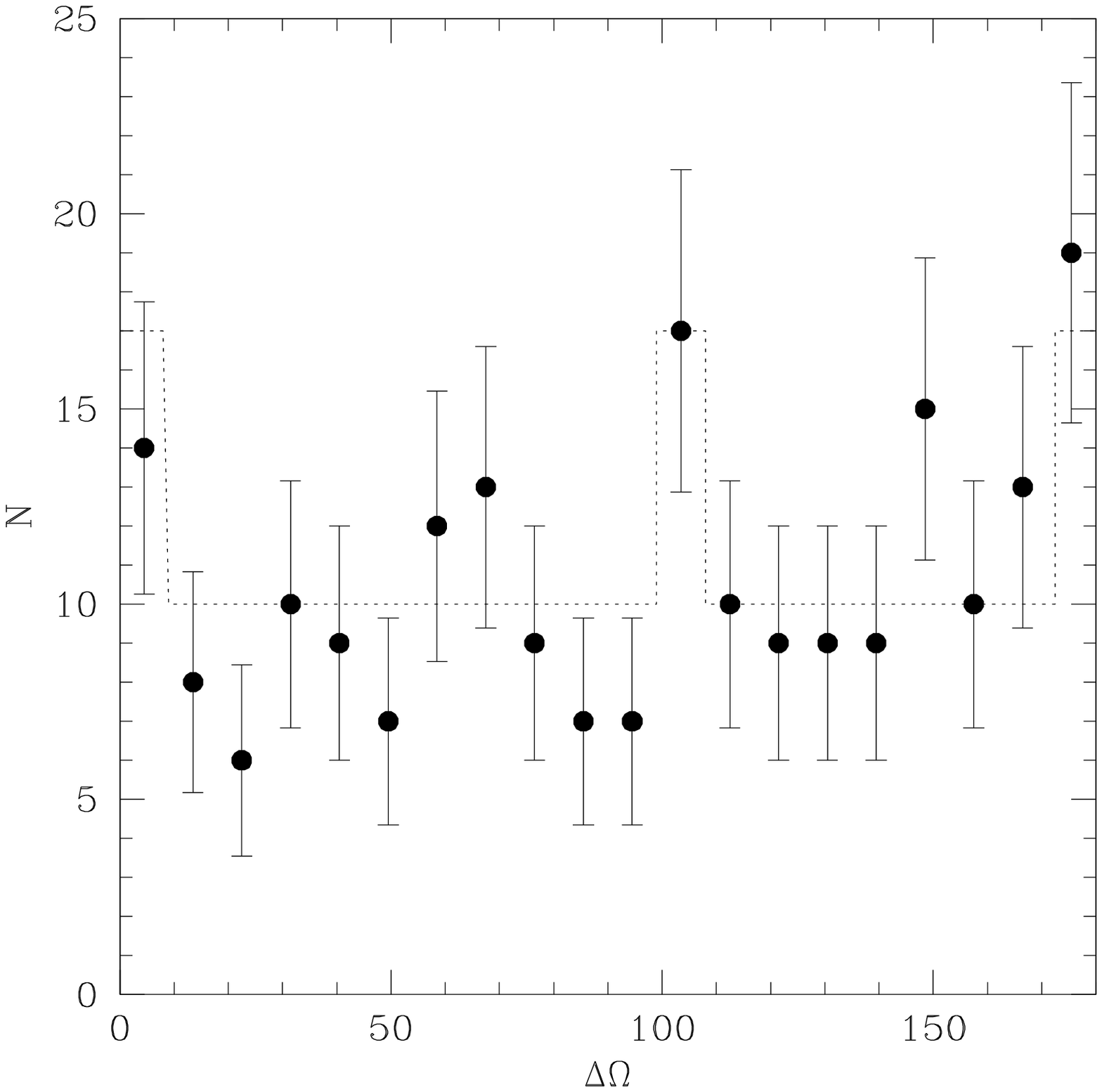}
\figcaption[DeloM.ps]{The points show the distribution of $\Delta \Omega$, which is the angle between the two lines of
nodes of a pair of neighbouring planets. The dotted line indicates the model distribution we use, which is flat except
for modest enhancements at a handful of preferred values.
\label{DeloM}}

\clearpage
\plotone{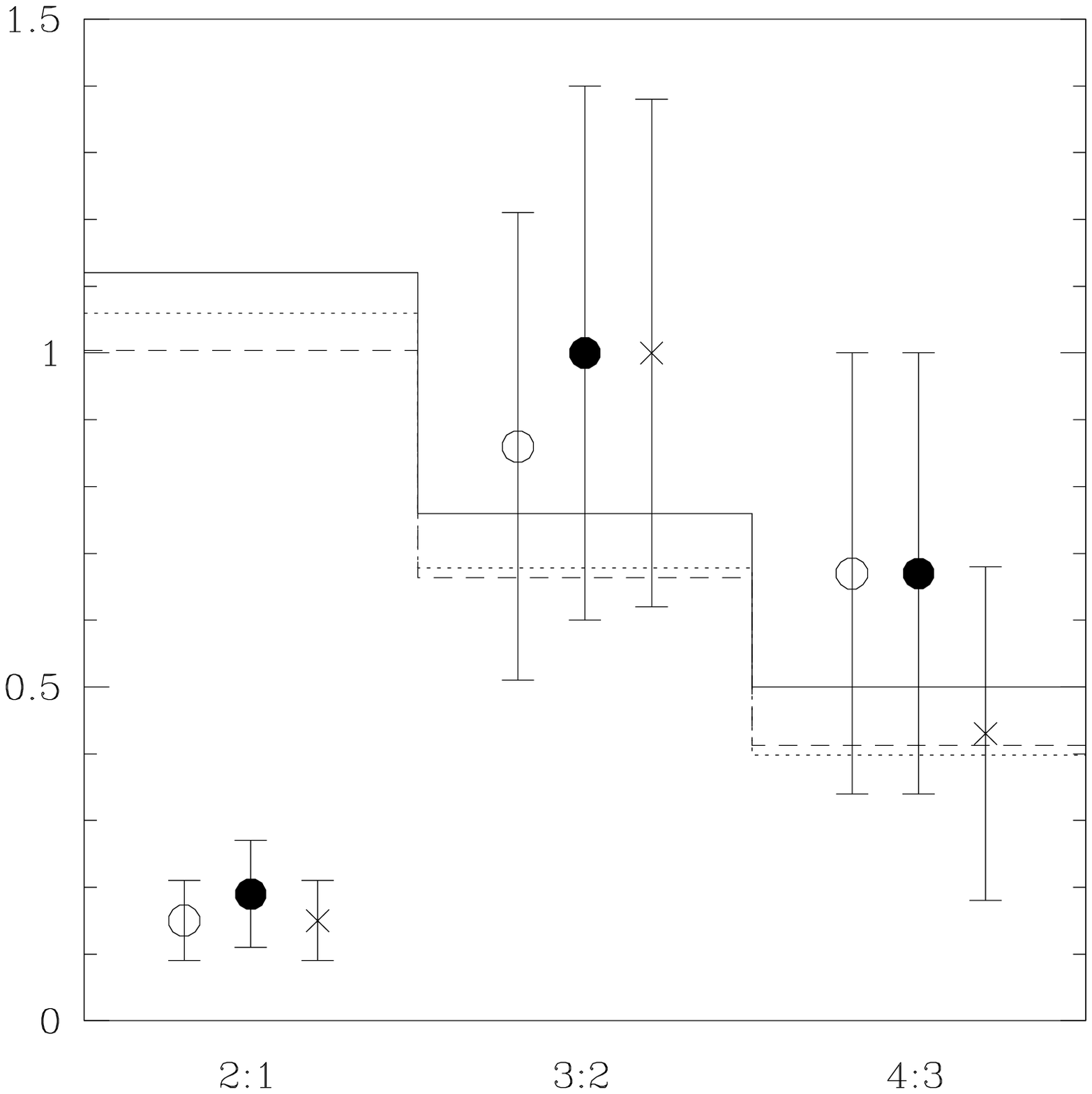}
\figcaption[MultiComp.ps]{The solid points indicate the multiplicity ratios calculated for the sample
identified in the dotted box in Figure~\ref{PRcomp}. The open circles represent the same quantity calculated
using the entire sample. The crosses indicate the multiplicities using the catalog of Muirhead et al. (2012).
The solid histogram represents our default model, with $R'=1.4$, and the dotted line uses $R'=1.2$. The
dashed line uses $R'=1.4$ for $a>0.07$AU and $R'=1$ otherwise. We see that the agreement is robust to
both variation in the observational catalogue and in the variation of model parameters.
\label{MultiComp}}

\clearpage
\plotone{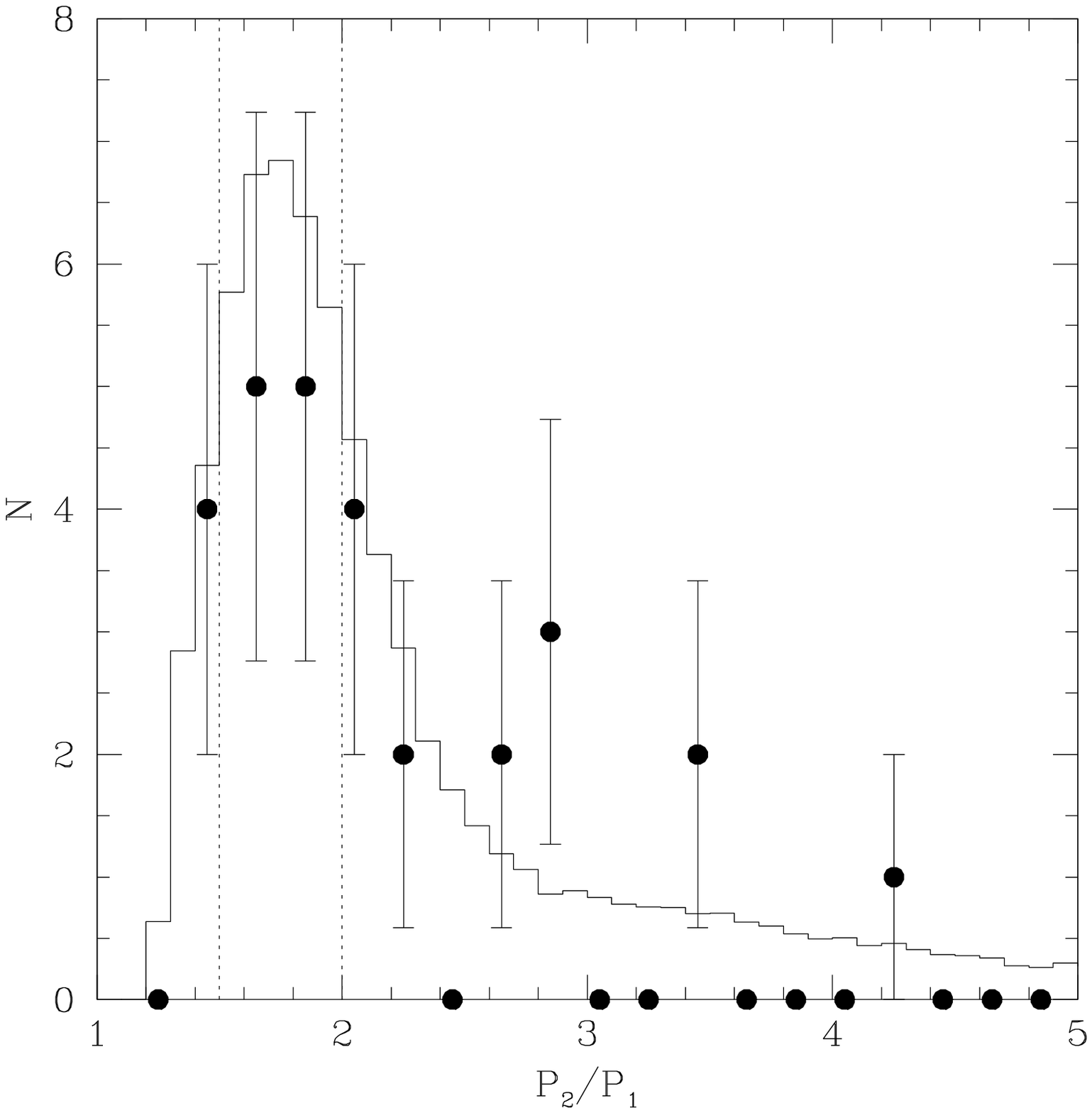}
\figcaption[PratM.ps]{The filled circles indicate the binned distribution of
period ratios for neighbouring tranets in the DC13 system. The histogram represents
the distribution that emerges from our Monte Carlo model with $R'=1.4$. The vertical
dotted lines indicate the 3:2 and 2:1 commensurabilities. We see that both the observations
and the simulations favour a broad distribution between 1.3 and 2.3.
\label{PratM}}

\clearpage
\plotone{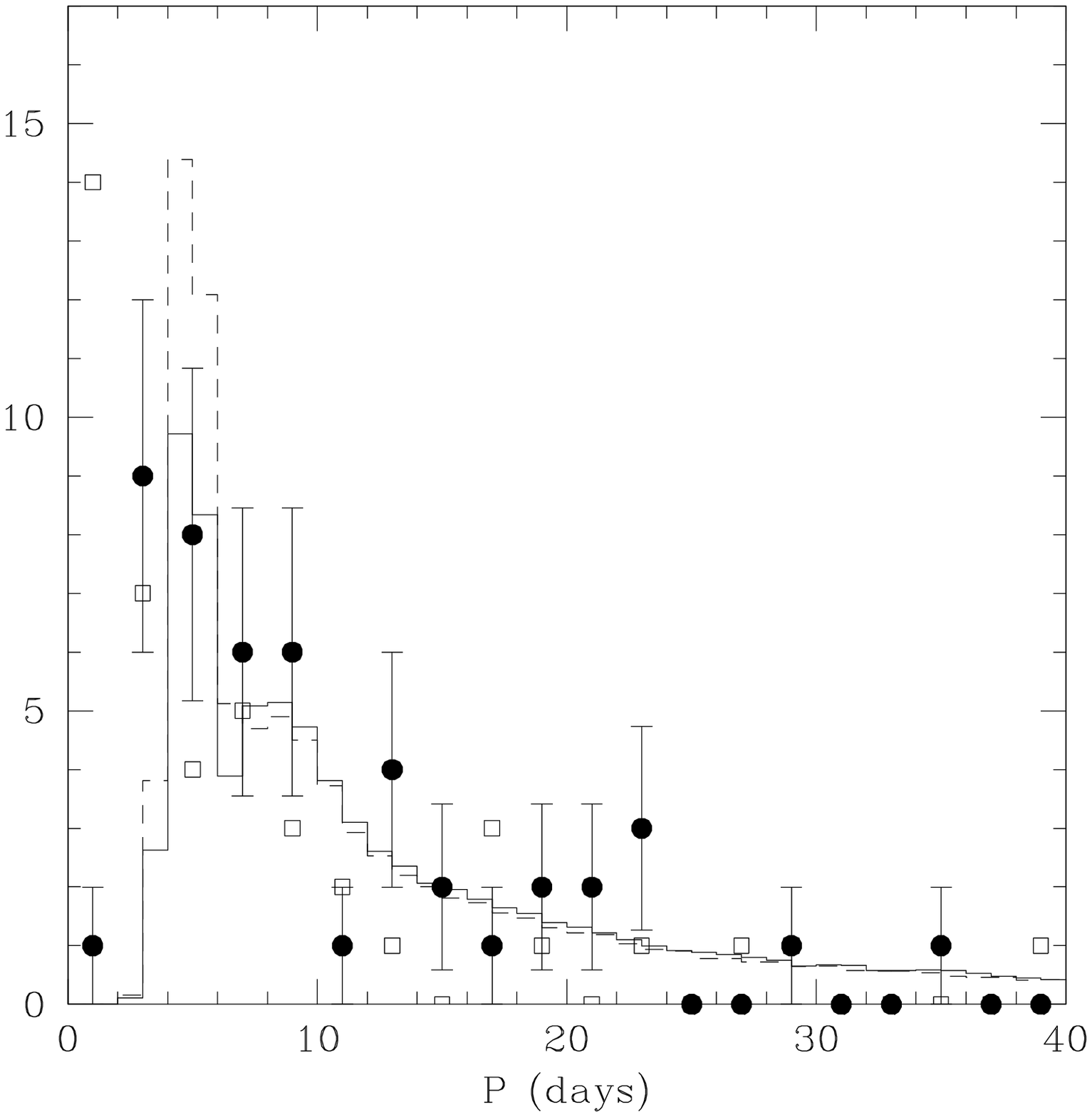}
\figcaption[PdisM.ps]{The filled circles show the binned period distribution of tranets in the DC13
catalogue that are found in multiple systems. The open squares shows the corresponding distribution
of single tranet systems. The excess of single tranets at periods $< 2$~days is evident. The dashed
histogram is the distribution from the Monte Carlo model with $R'=1.4$, while the solid histogram is
the distribution for the model in which $R'=1.4$ only for $a>0.07$~AU.
\label{Pdis}}

\clearpage
\plotone{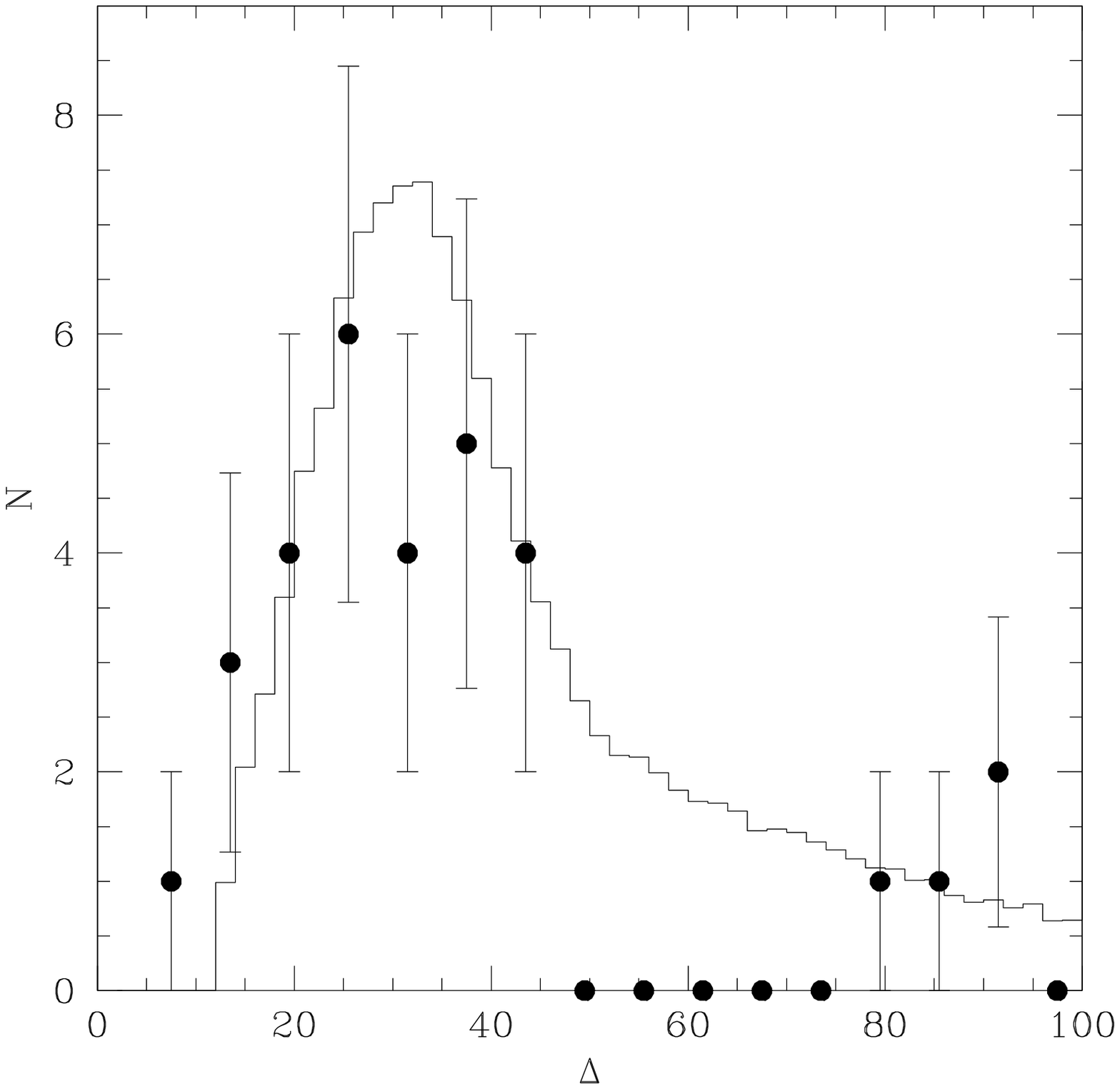}
\figcaption[DelM.ps]{The solid points are the observed values of $\Delta$, derived by calculating
$\Delta_L$ (using the Lissauer et al. 2011 relation) and then multiplying by factor of 1.5. The
histogram results from the Monte Carlo model, with $R'=1.4$, although the shape of the distribution
is rather insensitive to the value of $R'$.
\label{DelM}}

\clearpage
\plotone{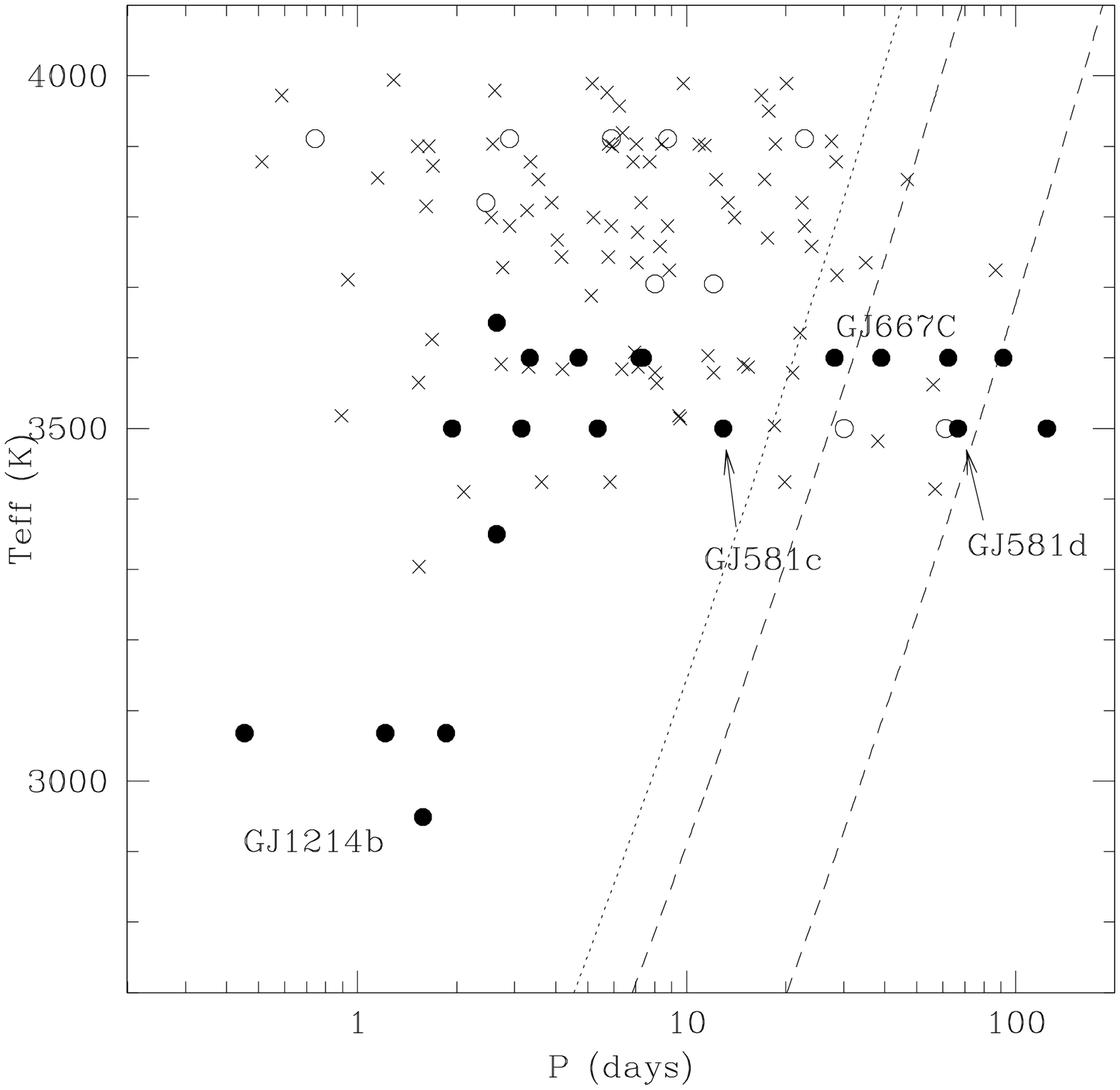}
\figcaption[Teqp2.ps]{Filled and open circles represent confirmed planets, showing orbital periods as
a function of host star effective temperature. The filled circles are planets whose minimum mass are
$< 0.1 M_J$, while open circles have masses $>0.1 M_J$. The crosses indicate planetary
candidates from the sample of DC13. 
 The dashed curves show the expectation of the 
habitable zone inner and outer edges from Kopparapu et al. (2013), assuming the `runaway greenhouse'
and `early Mars' models. The dotted line represents the more optimistic `recent Venus' model.
The listed names indicate several well-known M-dwarf planets, including
those most likely to be considered habitable.
\label{Teqp}}

\clearpage
\plotone{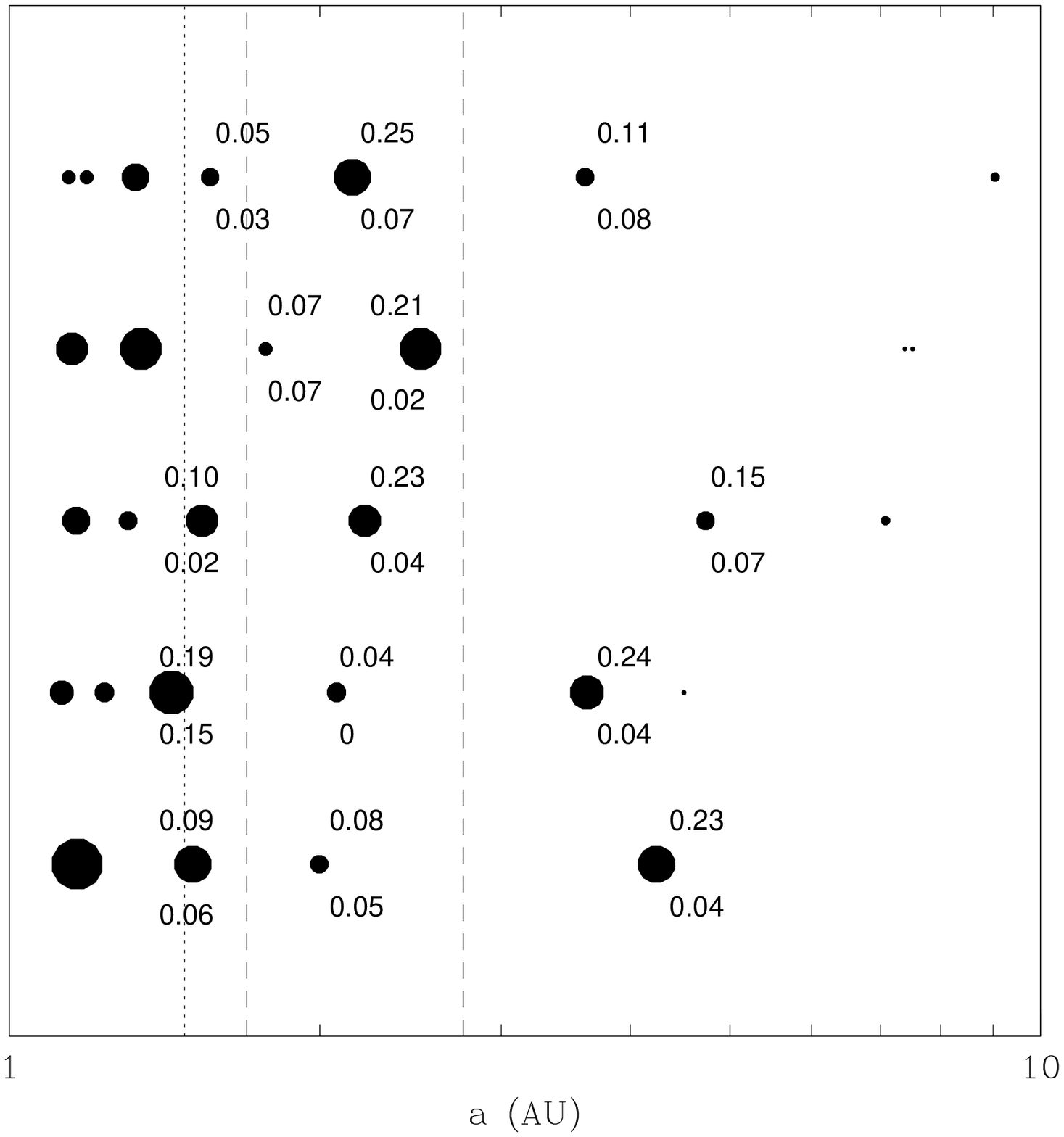}
\figcaption[Water.ps]{
Filled circles indicate planets at the end of our accumulation runs, in five different realisations
of the same model. The circle diameters scale linearly with planetary mass, to better illustrate the
mass variation. The upper labels indicate the fraction of all test particles in the range 0.5--1~AU
accreted by this body. The lower labels indicate only that fraction accreted after the last giant impact.
The vertical dashed lines indicate the inner and outer regions of the conservative habitable zone, while
the dotted line indicates the location of a more optimistic estimate of the inner edge. These criteria
are drawn from the models of Koppurapu et al. (2013) assume a 0.5$ M_{\odot}$ host.
\label{Water}}

\clearpage
\plotone{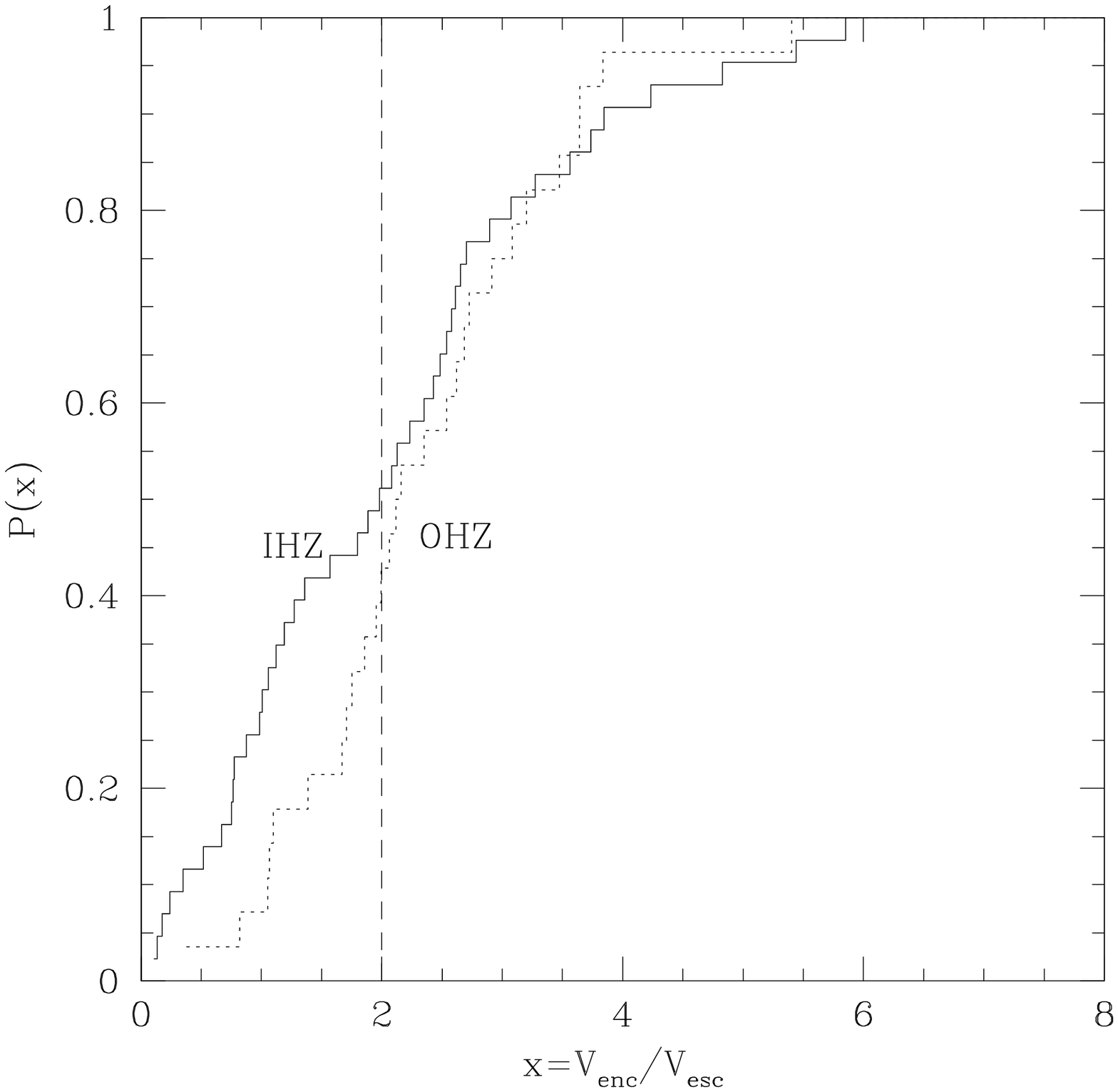}
\figcaption[Vesc.ps]{The solid histogram, labelled IHZ, shows the distribution of escape velocity-normalised
encounter velocities for test particles accreted by planets that lie between 0.17--0.27~AU in our simulations. Only
those particles accreted after the last major impact are counted. The dotted histogram represents the same
distribution for those planets that lie within 0.27--0.38~AU. The vertical dashed line indicates a value
of $V_{encounter}/V_{escape}=2$, which is the threshold for significant atmosphere erosion (Melosh \& Vickery 1989).
\label{Vesc}}

\clearpage
\plotone{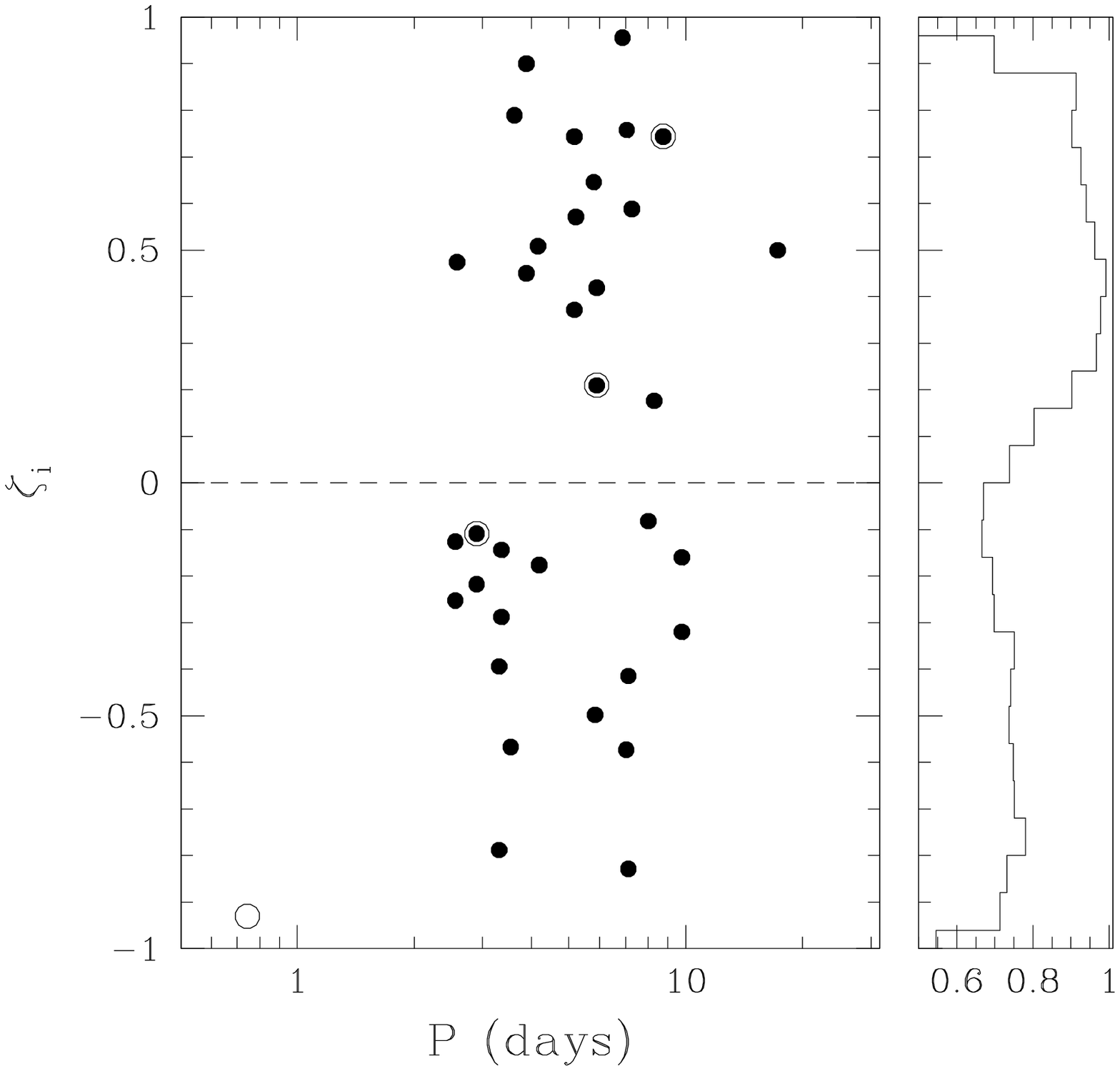}
\figcaption[Zeta.ps]{Filled circles show the values of $\zeta_i$ for the planet pairs in the DC13 sample. Values
of $\zeta_1$ are plotted unless $\left|\zeta_1\right|>1$, in which case $\zeta_2$ is plotted. The open circles
indicate the period ratios of the Kepler-32 system, which are included in the DC13 sample except for Kepler-32f (Swift et al. 2013),
which yields the open circle on the bottom left. The right hand panel shows the distribution of $\zeta_1$ that
emerges from the Monte Carlo model. Note the zoomed-in scale -- the distribution is broadly uniform in $\zeta$.
\label{Zeta}}

\clearpage
\plotone{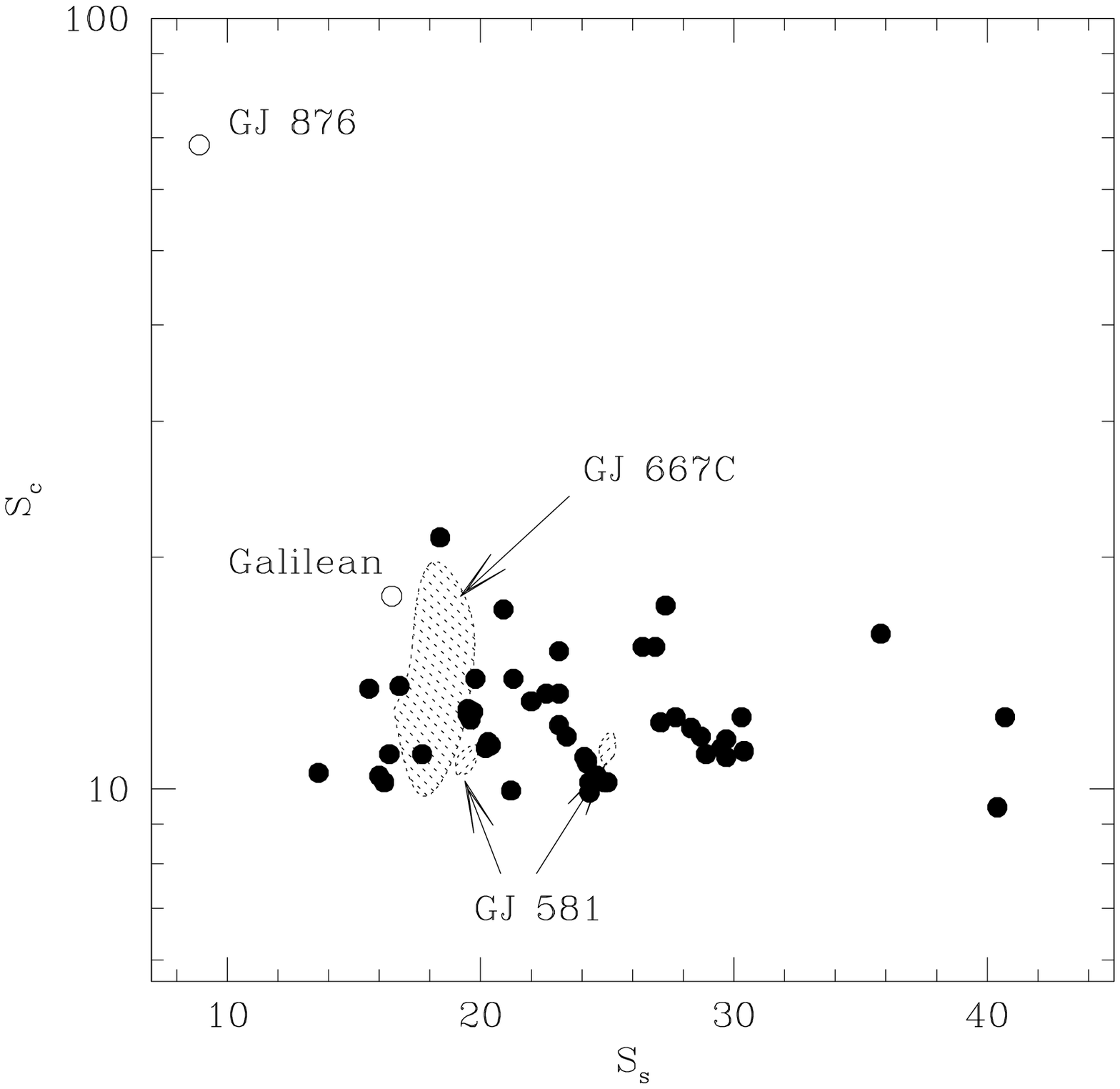}
\figcaption[Stat2.ps]{The filled circles show the statistical measures that characterise the simulations, given in
Table~\ref{OutTab}. The quantity $S_s$ measures the average separation between planets in terms of their mutual
Hill radii, while $S_c$ is a measure of how evenly spread the system mass is between the various components
(low values imply an even spread, and high values imply a concentration into only some members). The shaded regions
indicate the estimates values for the GJ~667C and GJ~581 systems, accounting for the $1\sigma$ uncertainties in
the masses. We show two regions for GJ~581, corresponding to whether we include GJ~581g (lower $S_s$ value) or not.
The open circles show two other systems of potential applicability -- the Galilean moons of Jupiter and the four-planet
system around the M-dwarf GJ~876. However, in both cases there is room to question the applicability of the model.
The Galilean moons have similar mass ratios, but the evident commensurabilities are usually taken as evidence of 
migration from their formation positions. The host star GJ~876 is indeed an M-dwarf, but two of the planets are
of Jovian mass, implying substantial gas accretion not accounted for in this model. 
\label{Stat2}}

\begin{deluxetable}{lccccccc}
\tablecolumns{8}
\tablewidth{0pc}
\tablecaption{Final States of Assembly Simulations: 
\label{OutTab}}
\tablehead{ Simulation &
\colhead{N} & \colhead{$M_{big}$}   & \colhead{$<a>_M$}  & \colhead{$S_d$} & \colhead{S$_s$} & \colhead{S$_c$} & \colhead{N$_{HZ}$} \\
 & \colhead{($M_{\oplus}$)} & \colhead{(AU)} &  &\colhead{($\times 10^{-2}$)} & & \\
 }
\startdata

  1 &  5 &    2.0 &   0.40 & 0.20     & 24.3     & 10.2     &  2 \\
   
  2 &  7 &    1.5 &   0.34 & 0.19     & 17.7     & 11.1     &  2 \\
   
  3 &  4 &    1.7 &   0.46 & 0.19     & 28.3     & 12.0     &  2 \\
   
  4 &  5 &    2.0 &   0.60 & 0.20     & 24.9     & 10.2     &  2 \\
   
  5 &  6 &    1.5 &   0.33 & 0.18     & 19.7     & 12.6     &  2 \\
   
  6 &  5 &    1.7 &   0.76 & 0.22     & 24.3     & 9.90     &  1\\
   
  7 &  5 &    1.7 &   0.31 & 0.17     & 22.6     & 13.3     &  3\\
   
  8 &  6 &    1.5 &   0.38 & 0.19     & 20.4     & 11.4     &  2\\
   
  9 &  5 &    1.2 &   0.37 & 0.17     & 23.1     & 13.3     &  2\\
   
 10 &  4 &    1.7 &   0.86 & 0.17     & 26.9     & 15.3     &  1\\
   
 11 &  6 &    1.5 &   0.30 & 0.15     & 19.8     & 13.9     &  3\\
   
 12 &  5 &    1.7 &   0.61 & 0.19     & 23.4     & 11.7     &  1\\
   
 13 &  8 &    1.3 &   0.12 & 0.19     & 16.2     & 10.2     &  4\\
   
 14 &  6 &    1.6 &   0.25 & 0.19     & 21.2     & 9.95     &  1\\
   
 15 &  5 &    1.8 &   0.24 & 0.16     & 20.9     & 17.1     &  1\\
   
 16 &  4 &    2.3 &   0.22 & 0.18     & 29.5     & 11.3     &  1\\
   
 17 &  4 &    2.2 &   0.58 & 0.17     & 27.1     & 12.2     &  2\\
   
 18 &  6 &    1.2 &   0.09 & 0.20     & 20.3     & 11.5     &  2\\
   
 19 &  4 &    2.0 &   0.28 & 0.20     & 29.7     & 11.0     &  1\\
   
 20 &  4 &    2.0 &   1.66 & 0.18     & 28.9     & 11.1     &  2\\
   
 21 &  4 &    2.5 &   0.88 & 0.20     & 30.3     & 12.4     &  1\\
   
 22 &  4 &    1.7 &   0.43 & 0.18     & 29.7     & 11.6     &  2\\
   
 23 &  6 &    1.2 &   0.29 & 0.18     & 19.5     & 12.7     &  3\\
   
 24 &  6 &    1.3 &   0.42 & 0.21     & 20.2     & 11.3     &  2\\
   
 25 &  5 &    1.8 &   0.37 & 0.19     & 22.0     & 13.0     &  2\\
   
 26 &  5 &    1.8 &   0.37 & 0.19     & 22.0     & 13.0     &  2\\
   
 27 &  6 &    1.3 &   0.11 & 0.18     & 19.5     & 12.5     &  2\\
   
 28 &  6 &    2.1 &   0.44 & 0.20     & 19.6     & 12.3     &  2\\
   
 29 &  5 &    2.1 &   0.37 & 0.10     & 18.4     & 21.2     &  1\\
   
 30 &  5 &    1.8 &   0.64 & 0.18     & 24.1     & 11.0     &  1\\
   
 31 &  5 &    1.5 &   0.52 & 0.20     & 24.2     & 10.8     &  1\\
   
 32 &  8 &    1.3 &   0.28 & 0.18     & 16.4     & 11.1     &  2\\
   
 33 &  4 &    1.8 &   0.59 & 0.15     & 26.4     & 15.3     &  1\\
   
 34 &  4 &    2.1 &   0.46 & 0.15     & 27.3     & 17.3     &  1\\
   
 35 &  8 &    1.0 &   0.11 & 0.18     & 15.6     & 13.5     &  3\\
   
 36 &  5 &    1.8 &   0.34 & 0.20     & 24.2     & 10.9     &  1\\
   
 37 &  5 &    2.0 &   0.47 & 0.20     & 24.6     & 10.4     &  1\\
   
 38 & 10 &    1.7 &   0.11 & 0.21     & 13.6     & 10.5     &  3\\
   
 39 &  4 &    2.5 &   0.25 & 0.18     & 30.4     & 11.2     &  1\\
   
 40 &  4 &    2.2 &   0.65 & 0.18     & 28.7     & 11.7     &  1\\
   
 41 &  7 &    1.5 &   0.14 & 0.18     & 16.8     & 13.6     &  3\\
   
 42 &  5 &    1.5 &   0.51 & 0.17     & 23.1     & 15.1     &  2\\
   
 43 &  8 &    1.5 &   0.15 & 0.21     & 16.0     & 10.4     &  2\\
   
 44 &  5 &    2.2 &   0.17 & 0.19     & 23.1     & 12.1     &  1\\
   
 45 &  3 &    2.2 &   1.15 & 0.22     & 40.4     & 9.47     &  1\\
   
 46 &  4 &    2.0 &   0.24 & 0.17     & 27.7     & 12.4     &  1\\
   
 47 &  3 &    2.0 &   1.16 & 0.16     & 35.8     & 15.9     &  1\\
   
 48 &  6 &    2.7 &   0.22 & 0.17     & 21.3     & 13.9     &  2\\
   
 49 &  3 &    2.7 &   1.81 & 0.17     & 40.7     & 12.4     &  1\\
   
 50 &  5 &    2.0 &   0.50 & 0.18     & 25.0     & 10.2     &  2\\
\enddata
\tablenotetext{}{The columns show the number $N$ of final planets with semi-major
axis $<0.5$~AU, the mass of the largest planet, and the four statistical measures $<a>_M$ (mass weighted semi-major axis),
$S_d$ (angular momentum deficit -- a measure of deviation from circular, coplanar orbits), $S_s$ (a measure of how closely
spaced the planets are in terms of gravitational influence) and $S_c$ (a measure of how bunched/spread out the mass distribution is).
The final column shows N$_{HZ}$, the number of surviving planets with semi-major axis between 0.17--0.38~AU (our nominal habitable zone).}
\end{deluxetable}

\begin{references}
\reference{Ali} Alibert, Y. et al., 2006, A\&A, 455, L25
\reference{ARKW} Andrews, S. M., Rosenfeld, K. A., Kraus, A. L. \& Wilner, D. J., 2013, arXiv:1305.5262
\reference{AE1} Anglada-Escude, G. et al., 2012, ApJ, 751, L16
\reference{AE2} Anglada-Escude, G. et al., 2013, arXiv:1306.6074
\reference{Barc} Barclay, T. et al., 2013, ApJ, 768, 101 
\reference{Bat12} Batalha, N. et al., 2013, ApJS, 204, 24
\reference{Bon11} Bonfils, X. et al., 2011, arXiv:1111.5019
\reference{Bor11} Borucki, W. J., et al., 2011, ApJ, 736, 19
\reference{Bor13} Borucki, W. J. et al., 2013, Science, 340, 587
\reference{C99} Chambers, J. E., 1999, MNRAS, 304, 793
\reference{CL} Chiang, E. \& Laughlin, G., 2013, MNRAS, 431, 3444
\reference{GS14} Goldreich, P. \& Schlichting, H. E., 2014, AJ, 147, 32
\reference{Del} Delfosse, X., et al., 2013, A\&A, 553, 8
\reference{DC2} Drake, M. J. \& Campins, H., 2006, Ast, Comet \& Meteors, 229, 381
\reference{DC} Dressing, C. D. \& Charbonneau, D., 2013, ApJ, 767, 95
\reference{F12} Fabrycky, D. et al., 2012, arXiv:1202.6328 
\reference{F11} Forveille, T. et al., 2011, arXiv:1109.2505
\reference{G10} Gould, A., et al., 2010, ApJ, 720, 1073
\reference{H10} Haghighipour, N. et al., 2010, ApJ, 715, 271
\reference{H13}  Haghighipour, N., 2013, Ann. Rev. Earth. Planet. Sci., 2013, 41, 469
\reference{H09} Hansen, B., 2009, ApJ, 703, 1131
\reference{Rocks} Hansen, B. \& Murray, N., 2012, ApJ 751, 158
\reference{RocksII} Hansen, B. \& Murray, N., 2013, arXiv:1301.7431
\reference{How10} Howard, A. et al., 2010, Science, 330, 51
\reference{How} Howard, A. et al., 2012, ApJS, 201, 15
\reference{HA} Hughes, A. L. \& Armitage, P. J., 2012, MNRAS, 423, 389
\reference{IdL8} Ida, S. \& Lin, D. N. C., 2008, ApJ, 685, 84
\reference{IdL10} Ida, S. \& Lin, D. N. C., 2010, ApJ, 719, 810
\reference{Izo} Izidoro, A., de Souza Torres, K., Winter, O. C. \& Haghighipour, N., 2013, ApJ, 767, 54
\reference{J09} Johnson, J. A., Butler, R. P., Marcy, G. W., Fischer, D. A., 
Vogt, S. S., Wright, J. T. \& Peek, K. M. G., 2007, ApJ, 670, 833
\reference{J12} Johnson, J. A. et al., 2012, AJ, 143, 111
\reference{J97} Joshi, M. M., Haberle, R. M. \& Reynolds, R. T., 1997, Icarus, 129, 450
\reference{Kas93} Kasting, J. F.,Whitmire, D. P. \& Reynolds, R. T., 1993, Icarus, 101, 108
\reference{KI} Kokubo, E. \& Ida, S., 1998, Icarus, 131, 171
\reference{K131} Kopparapu, R. K. et al., 2013, ApJ, 765, 131
\reference{K132} Kopparapu, R. K., 2013, ApJ, 767, L8
\reference{L13} Lammer, H. et al., 2013, MNRAS, 430, 1247
\reference{L97} Laskar, J., 1997, A\&A, 317, L75
\reference{LBA} Laughlin, G., Bodenheimer, P. \& Adams, F., 2004, ApJ, 612, L73
\reference{L07} Lissauer, J. J., 2007, ApJ, 660, L149
\reference{L11} Lissauer, J. J. et al., 2011, ApJS, 197, 8
\reference{LFM} Lopez, E. D., Fortney, J. J. \& Miller, N., 2012, ApJ, 761, 59
\reference{M12} Mann, A. W., Gaidos, E., Lepine, S. \& Hilton, E. J., 2012, ApJ, 753, 90
\reference{Marty} Marty, B., 2012, Earth \& Plan. Sci. Letters, 313, 56
\reference{M09b} Mayor, M. et al., 2009b, A\&A, 507, 487
\reference{M11} Mayor, M., et al., 2011, arXiv:1109.2497
\reference{MV} Melosh, H. J. \& Vickery, A. M., 1989, Nature, 338, 487
\reference{M2000} Morbidelli, A., Chambers, J., Lunine, J. I., Petit, J. M., Robert, F.,
Valsecchi, G. B. \& Cyr, K. E., 2000, M\&PS, 35, 1309
\reference{Mord} Mordasini, C., Alibert, Y. \& Benz, W., 2009, A\&A, 501, 1139
\reference{ML09} Montgomery, R. \& Laughlin, G., 2009, Icarus, 202, 1
\reference{MS13} Morton, T. D. \& Swift, J., 2013, arXiv:1303.3013
\reference{Muir} Muirhead, P. S., Hamren, K., Schlawin, E., Rojas-Ayala, B., Covey, K. \& Lloyd, J. P., 2012, ApJ, 750, L37
\reference{OI} Ogihara, M. \& Ida, S., 2009, ApJ, 699, 824
\reference{OJ} Owen, J. E. \& Jackson, A. P., 2012, MNRAS, 425, 2931
\reference{OW} Owen, J. E. \& Wu, Y., 2013, arXiv:1303.3899
\reference{Ray07} Raymond, S. N., Scalo, J. \& Meadows, V. S., 2007, ApJ, 669, 606
\reference{Ray08} Raymond, S. N., Barnes, R. \& Mandell, A. M., 2008, MNRAS, 384, 663
\reference{R12} Rein, H., 2012, MNRAS, 427, L21
\reference{R05} Rivera, E. J., Laughlin, G., Butler, R. P., Vogt, S. S., Haghighipour, N. \& Meschiari, S., 2010, ApJ, 719, 890
\reference{S13} Swift, J. J., Johnson, J. A., Morton, T. D., Crepp, J. R., Montet, B. T., 
Fabrycky, D. C. \& Muirhead, P. S., 2013, ApJ, 764, 105
\reference{TP07} Terquem, C. \& Papaloizou, J. C. B., 2007, ApJ, 654, 1110
\reference{TD} Tremaine, S. \& Dong, S., 2012, AJ, 143, 94
\reference{Ud} Udry, S. et al., 2007, A\&A, 469, L43
\reference{Vog} Vogt, S. et al., 2010, ApJ, 723, 954
\reference{VBH} Vogt, S. S., Butler, R. P. \& Haghighipour, N., 2012, AN, 333, 561
\reference{Weid} Weidenschilling, S. J., 1977, Ap\&SS, 51, 153
\reference{Youd} Youdin, A., 2011, ApJ, 742, 38
\end{references}
\end{document}